%% file: SecondSolar.tex
\PassOptionsToPackage{pdfpagelabels=false}{hyperref} 
\documentclass[a4paper,fleqn,usenatbib]{mnras}
\usepackage{mathptmx}
\usepackage{pdflscape}
\usepackage[T1]{fontenc}
\usepackage{ae,aecompl}
\usepackage{graphicx}	
\usepackage{amssymb}	
\usepackage{amsmath}	
\usepackage{hhline}

\title[The polarised spectrum of 89\,Her]{The solar-like "Second Spectrum" and the polarised metal lines in emission of the post-AGB binary 89\,Herculis.}

\author[F. Leone et al.]{F. Leone$^{1,2},$
M. Gangi$^{1,2}$,
M. Giarrusso$^{1,2}$,
C. Scalia$^{1,2}$, 
M. Cecconi$^{3}$,  
R. Cosentino$^{3}$,\and
A. Ghedina$^{3}$,
M. Munari$^{2}$,
S. Scuderi$^{2}$
\\
$^{1}$Universit\`a di Catania, Dipartimento di Fisica e Astronomia, Sezione Astrofisica, Via S. Sofia 78, I--95123 Catania, Italy \\
$^{2}$INAF - Osservatorio Astrofisico di Catania, Via S. Sofia 78, I--95123 Catania, Italy\\
$^{3}$INAF - Fund. Galileo Galilei, Rambla Jos\'e Ana Fern\'andez Perez 7, 38712 Bre\~na Baja (La Palma), Canary Islands, Spain
}

\date{Accepted XXX. Received YYY; in original form ZZZ}

\pubyear{2016}

\begin{document}
\label{firstpage}
\pagerange{\pageref{firstpage}--\pageref{lastpage}}
\maketitle

\begin{abstract}
We studied the polarised spectrum of the post-AGB binary system 89\,Herculis on the basis of data collected with the high resolution
\emph{Catania Astrophysical Observatory Spectropolarimeter}, \emph{HArps-North POlarimeter}  and  \emph{Echelle SpectroPolarimetric Device 
for the Observation of Stars}. We find the existence of linear polarisation in the strongest metal  lines in absorption and with low excitation potentials.
Signals are characterized by complex Q and U morphologies varying with the orbital period. As possible origin of this "Second Solar Spectrum"-like behaviour, we rule out
magnetic fields, continuum depolarisation due pulsations and hot spots. { The linear polarisation we detected also in the Ca{\sc ii}\,8662\AA\, line is a clear
evidence of optical pumping polarisation and it rules out the scattering polarisation from free electrons of the circumbinary environment.}
In the framework of optical pumping due to the secondary star, the observed periodic properties of the spectral line polarisation can be justied by two jets, 
flow velocity of few tens of km\,s$^{-1}$, at the basis of that hour-glass structure characterising 89\,Herculis. We also discovered linear polarisation across the emission profile of metal lines. Numerical simulations show that these polarised profiles could be formed in an undisrupted circumbinary disk
rotating at $\le$10 km\,s$^{-1}$ and whose orientation in the sky is in agreement with optical and radio interferometric results.  We conclude that the study
of those aspherical enevlopes, whose origin is not yet completely understood, of PNe and already present in the post-AGB's, can benefit of high resolution
spectropolarimetry and that this technique can shape envelopes still too far for interferometry. 
\end{abstract}

\begin{keywords}
circumstellar matter - stars: post-AGB - techniques: polarimetric
\end{keywords}

\section{Introduction}
Photopolarimetry by \cite{1968AJ.....73..677K} has shown that the Mira
variable stars are characterised by a certain degree of linear polarisation changing with wavelength  and time. 
By means of spectropolarimetry at low resolution,  \cite{McLean1978} 
has pointed out a high polarisation degree, up to 7\%, across Balmer lines of Mira itself. \cite{Harrington2009, Harrington2009bis} and \cite{Lebre2011} have shown that such a property is often present in the Balmer lines of evolved stars.
The existence of linearly polarisation across individual metal lines in absorption has been reported for the Mira star $\chi$Cygni by \cite{Lebre2014a} and
for the red supergiant  Betelgeuse by \cite{Auriere2016}.
Using the Least Squares Deconvolution \citep[LSD,][]{Donati1997}, \cite{Lebre2015} found evidence of linear
polarisation in the spectral lines of the RV\,Tau variable R\,Scuti and \cite{Sabin2015} in the lines of the RV\,Tau variable U\,Monoceroti
as well in the post-AGB star 89\,Herculis. 

\begin{figure*}
\centering
\includegraphics[trim= 7cm 0.cm 3.0cm 1.0cm, clip=true,width=5.7cm,angle=180]{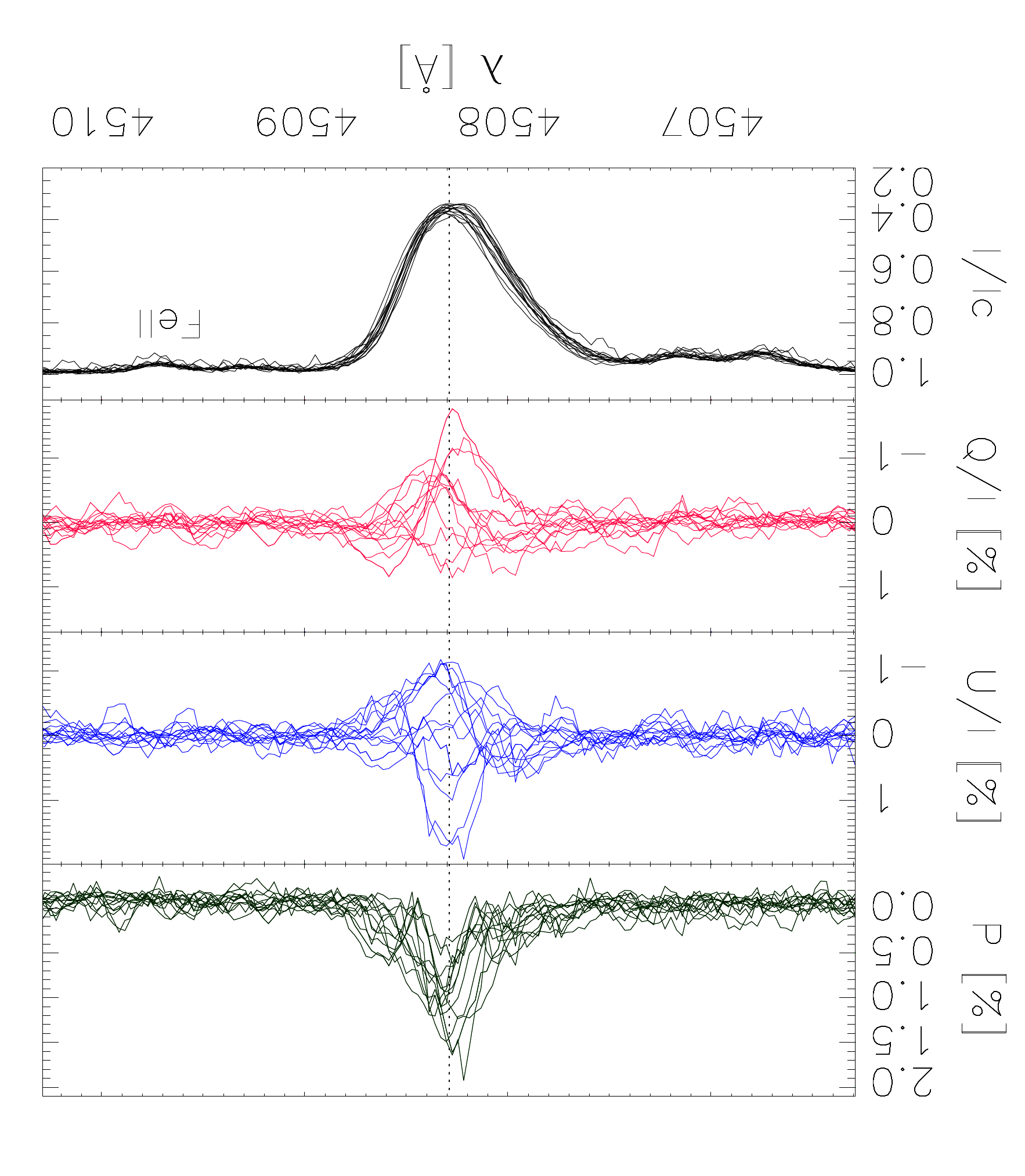}
\includegraphics[trim= 7cm 0.cm 3.0cm 1.0cm, clip=true,width=5.7cm,angle=180]{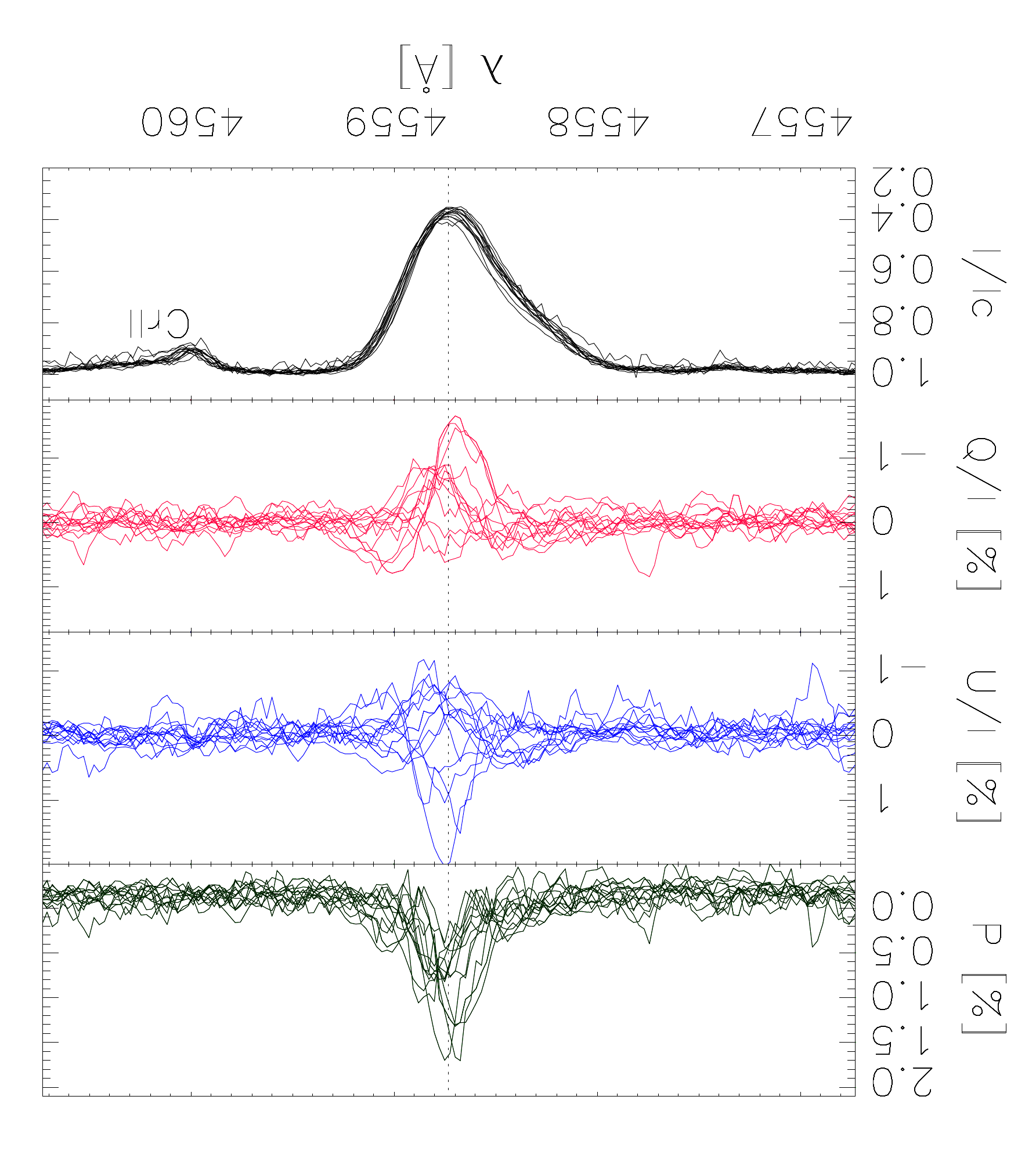}
\includegraphics[trim= 7cm 0.cm 3.0cm 1.0cm, clip=true,width=5.7cm,angle=180]{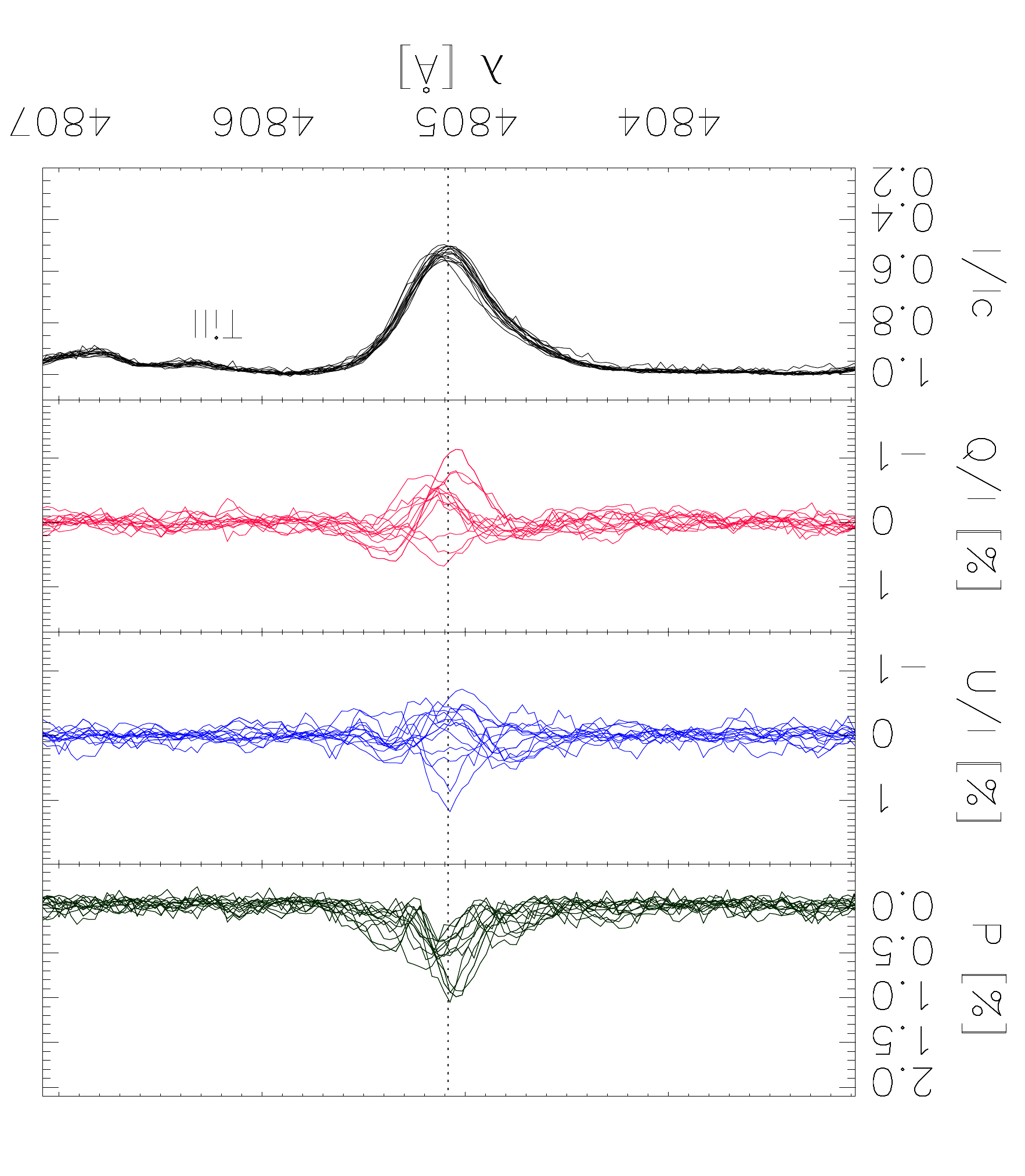}
\begin{center}
\caption{\label{Fig:PolExample} All 15 ESPaDOnS Stokes {\it I}, {\it Q/I}, {\it U/I} profiles and polarisation ($P=\int\sqrt{(Q/I)^2+(U/I)^2} d\lambda$) for the \ion{Fe}{II} 4508.288 \AA, \ion{Cr}{II} 4558.783 \AA \, and \ion{Ti}{II} 4805.085\,\AA\, lines. Data collected, between 2005 and 2009, present a clearly linear polarisation and a significant morphological variability in time.}
\end{center}
\end{figure*}

\begin{figure}
\centering
\includegraphics[trim= 0cm 0.3cm 7.0cm 15.0cm, clip=true,width=8.8cm]{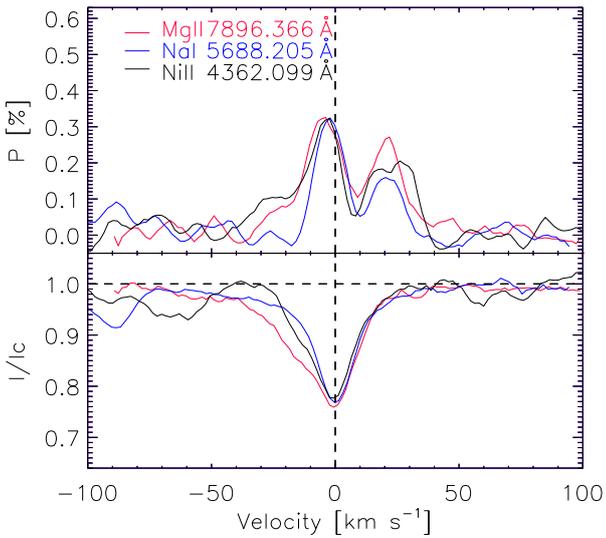}
\begin{center}\caption{\label{Fig:PolWave} Examples of spectral lines with equal depth showing that, within errors, polarisation does not depend on wavelength.}
\end{center}
\end{figure}

It appears that evolved stars can be characterised by the so called {\it Second Solar Spectrum}
consisting of spectral lines partially in absorption and partially in emission when observed
at high resolution linear spectropolarimetry \citep{Stenflo1982}. This spectrum in polarised light has represented
the observative basis for the comprehension of the most external layers of the Sun and now its discovery in evolved stars
opens new perspective for the understanding of stellar evolution. In general, polarisation across spectral lines is expected to trace and shape the enevlopes of the final stellar stages 
whose departure from the spherical symmetry is still a matter of discussion.
Despite many theories invoke magnetic fields to explain the rich variety of aspherical components observed in Planetary Nebulae (PNe)  \citep[see the review by][]{Balick2002},
no direct observational evidence of such fields has yet been established \citep{Leone2011, Jordan2012, Leone2014, Asensio2014, Sabin2015}. The presence of rotating circumbinary disks
is also a fundamental problem to understand the post-Asymptotic Giant Branch (post-AGB) evolution \citep{Buja2007}.
 
In this context, we report on one of the most known long term variables: 89\,Herculis (Section \ref{Sec:89Her}). A study based on high resolution
linear spectropolarimetry  (Section \ref{Sec:Observations}) carried out with the aim to point out any possible envelope asymmetry already present in the post-AGB phase before the PNe phase. In Section\,\ref{Sec:linepol}, we present the
\emph{Second Stellar Spectrum} of 89\,Her and its periodic variability. In Section \ref{Sec:Modeling}, we analyse the possible origins
of this polarisation. In Section \ref{Sec:Emission}, we present the polarisation of metal lines in emission from the circumbinary envelope
and relate their properties to the geometry and dynamic of the emitting region. Finally, we resume our conclusions in the last Section \ref{Sec:Discussions}.

\input{ObservingLog.tex}

\begin{figure*}
\centering
\includegraphics[trim= 9cm 2cm 3cm 2.5cm, clip=true,width=5.6cm,angle=180]{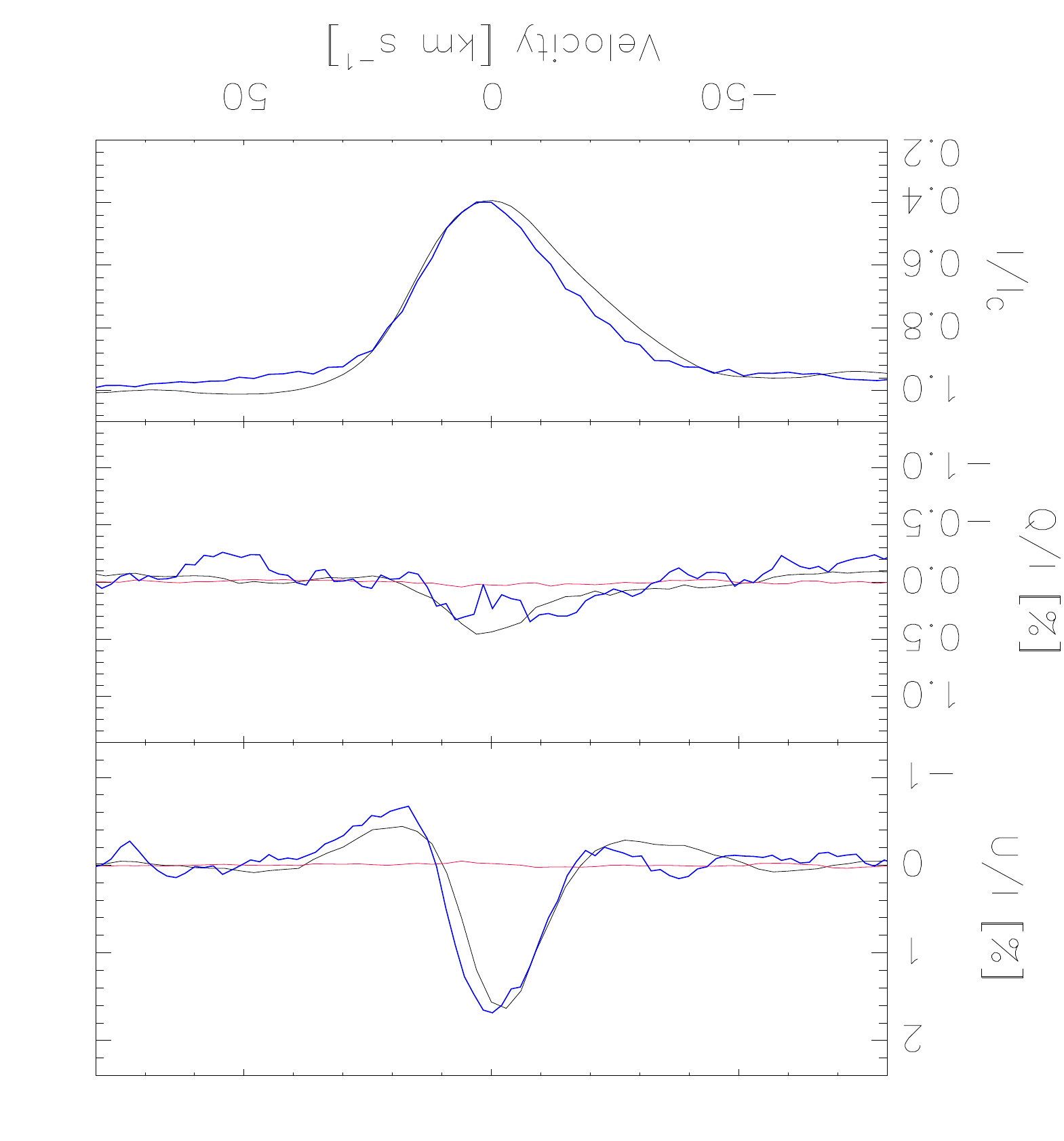}
\includegraphics[trim= 9cm 2cm 3cm 2.5cm, clip=true,width=5.6cm,angle=180]{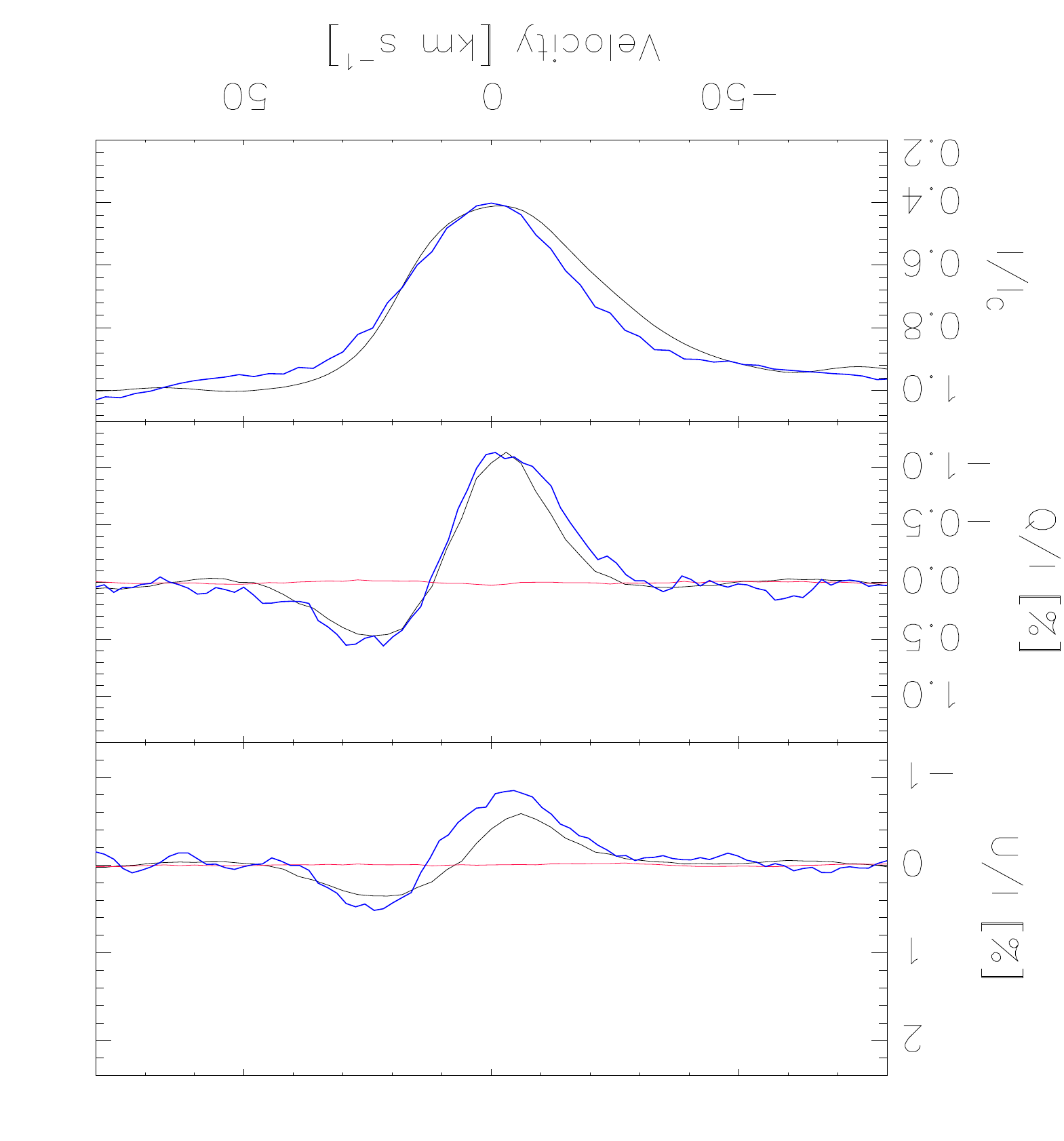}
\includegraphics[trim= 9cm 2cm 3cm 2.5cm, clip=true,width=5.6cm,angle=180]{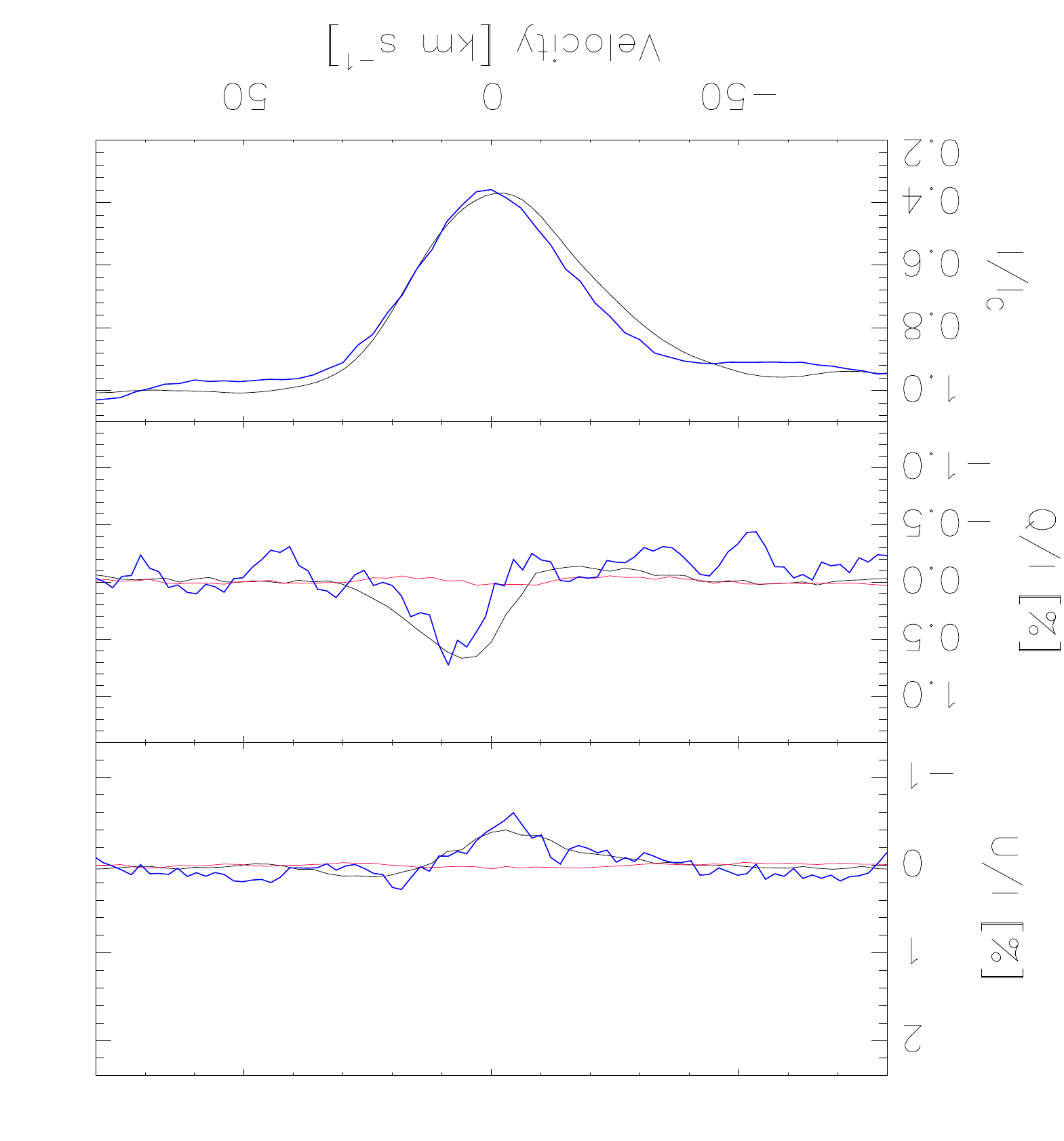}
\begin{center}\caption{\label{Fig:LSD} ESPaDOnS: Comparison between the Stokes parameters of the \ion{Fe}{II} $4508.288$ \AA \hspace{0.1cm}({\bf blue}) line
and LSD profiles ({\bf black}). LSD null polarisation in red. From the left, spectra were acquired on HJD\,53604.753, 53961.777 and 54878.132 data.}\end{center}
\end{figure*}

\section{89\,Herculis}\label{Sec:89Her}
Classified as an F2 Ib C supergiants by \cite{Gray1989}, 89\,Herculis (HR\,6685, HD\,163506) is the prototype of the new class, introduced by \cite{Waters1993}, of
post-AGB binaries surrounded by a circumbinary dust disk.  The ephemeris of this binary system was determined by Waters and coworkers from Radial Velocities (RV):
\begin{equation}
JD(\rm RV_{\rm max}) = 2\,446\,013.72(\pm 16.95)  + 288.36(\pm 0.71)\,\,\,\,\,\,\,\,\rm days\label{Ephemeris}
\end{equation}

Furthermore,  the primary component of 89\,Her pulsates with a period of 63.5 days \citep{Ferro1984}.

Interferometric data of 89\,Her have been interpreted by \cite{Buja2007}  with two nebular components: an expanding hour-glass
structure and an unresolved circumbinary Keplerian disk. Bujarrabal and coworkers concluded
that the hour-glass axis is tilted with respect to the Line-of-Sight (LoS) of $\theta_{LoS} = 15\degr$ with a Position Angle $\theta_{PA} = 45\degr$. 

From the basis of time resolved broad-band photopolarimetry, \cite{Akras2017} concluded that 89\,Her is an intrinsic unpolarised source in the visible range.

\section{Observational data}\label{Sec:Observations}
From June 2014 to October 2016, we have performed  high resolution linear spectropolarimetry of 89\,Her with the high resolution 
{\it Catania Astrophysical Observatory Spectropolarimeter}  \citep[CAOS, $R = 55\,000$,][]{Leone2016}  at the 0.91\,m telescope of
the stellar station of the Catania Astrophysical Observatory (\emph{G. M. Fracastoro} Stellar Station, Serra La Nave, Mt. Etna, Italy).
A spectrum in linearly polarised light was acquired on May 27, 2017 with the HArps-North POlarimeter \citep[HANPO, $R = 115\,000$,][]{LeoneH2016}
of the Telescopio Nazionale Galileo (Roque de Los Muchachos Astronomical Observatory, La Palma, Spain). 
Data were reduced and Stokes {\it Q/I}, {\it U/I} and \emph{null} spectra $N$ obtained  according to the procedures described in  \cite{Leone2016}.  The \emph{null} spectra check the presence of any spurious contribution to the polarised spectra and errors in the data reduction process. 

In addition, reduced spectropolarimetric data of 89\,Her have been retrieved from the Canadian Astronomy Data Centre. Spectra were collected from 2005 to 2009 at the $3.6$ $m$ Canadian-France-Hawaii Telescope with  the "Echelle SpectroPolarimetric Device for the Observation of Stars" \citep[ESPaDOnS, $R = 68\,000$,][]{Donati2006}. 

The logbook of observations is given in Table\,\ref{tab:ObsLog}. 

\section{Polarisation of metal lines in absorption}\label{Sec:linepol}
\citet{Sabin2015}  discovered evidence of linear polarisation in the Stokes {\it Q/I} and {\it U/I} LSD profiles of 89\,Her spectra recorded with ESPaDOnS on Feb. 8, 2006.
We found that all the archived ESPaDOnS spectra of 89\,Her, spanning about 1400 days,  are characterised by metal lines in absorption whose Stokes {\it Q/I} and {\it U/I} profiles
are not null and variable in time. Figure \ref{Fig:PolExample} provides an example of such variable signals for transitions of \ion{Fe}{II}, \ion{Cr}{II} and \ion{Ti}{II}. 
In any spectrum, Stokes profiles scale with the line depth irrespective of wavelength (Fig.\,\ref{Fig:PolWave}). We have then selected
in the 400-700 nm range, common to the three spectropolarimeters, a set of about 1000 metal lines whose Stokes $I$ residual
is larger then 0.2  to  obtain Stokes {\it Q/I} and {\it U/I} LSD profiles. 
Fig.\,\ref{Fig:LSD} shows the comparison between the ESPaDOnS-LSD and  \ion{Fe}{II} 4508.288\,\AA\, profiles at three different  dates.

\begin{figure}
\centering
\includegraphics[trim= 2.5cm 12.3cm 0.5cm 1.0cm, clip=true,width=15.5cm]{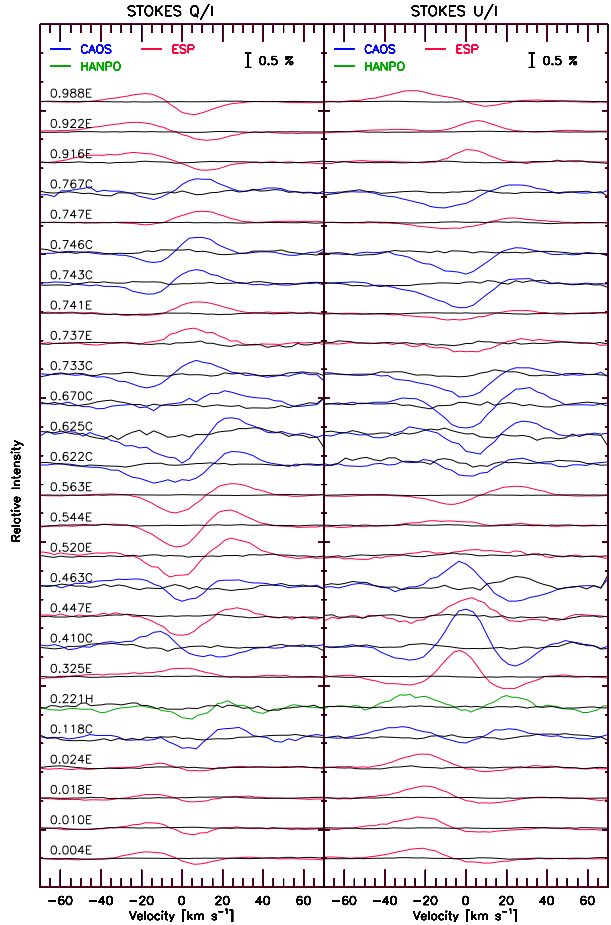}
\begin{center}\caption{\label{Fig:ResultsTime1} Observed Stokes {\it Q/I} and {\it U/I} LSD profiles of 89\,Her. Null polarisation
in black. Profiles are ordered according to the orbital phase computed with Eq.\,1 ephemeris
and arbitrarily shifted for a better visualisation. After the phase value, E means that the spectra were obtained with ESPaDOnS, C with CAOS and H with HANPO.}\end{center}
\end{figure}

In CAOS spectra of 89\,Her, we don't find any direct evidence of polarisation. After numerical simulations showing that ESPaDOnS Stokes {\it Q/I} and {\it U/I} profiles would be hidden in the
Signal-to-Noise (S/N) of CAOS spectra, we have successfully constructed  the CAOS-LSD profiles from the spectral line list we selected for the ESPaDOnS-LSD. For consistency, also the
HANPO-LSD Stokes profiles were based on the same list of spectral lines. The complete collection of the variable Stokes {\it Q/I} and {\it U/I} LSD profiles
is shown in Fig.\,\ref{Fig:ResultsTime1}.  S/N values achieved in \emph{null} LSD profiles is given in Table \ref{tab:ObsLog}. 

\begin{figure}
\centering\includegraphics[trim= 2.0cm 2.0cm 0.0cm 0.0cm, clip=true,width=8.2cm]{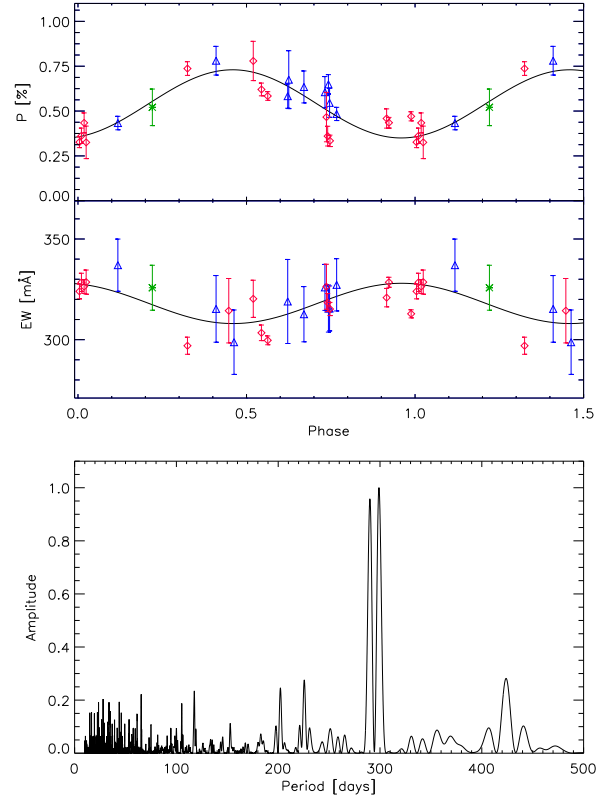}
\begin{center}
\caption{\label{Fig:Scargle} Bottom panel: CLEANED \citep{Roberts1987} Periodogram \citep{Scargle1982} of LSD polarisation of 89\,Her. The highest peak is close to the 288.36 day orbital period while
no power is present at the pulsation period of 63.5 days. Variations of the total polarisation $P$ (top panel) and Equivalent Width (EW, middle panel) of LSD profiles are plotted according to the orbital phase. sine fit is also plotted.
A sine fit of variability is also shown. Symbols are:
\textcolor{red}{$\diamond$}'s for ESPaDOnS data, \textcolor{blue}{\tiny $\triangle$}'s for CAOS data and \textcolor{green}{$*$}'s for HANPO data.}
\end{center}
\end{figure}

We have then performed a time analysis of the total polarisation measured across the LSD profiles. 
The \cite{Scargle1982} periodogram, CLEANED following \cite{Roberts1987},  presents
the highest peak at 294$\pm$5 days  that is coincident within errors with the orbital period (Fig.\,\ref{Fig:Scargle}). For this reason here after we will adopt the  ephemeris by \cite{Waters1993} (Eq.\,1).
Fig.\,\ref{Fig:Scargle} shows how  the total polarisation changes with the orbital period. A similar variability  cannot be statistically confirmed for the Equivalent Widths, whose sinusoidal fit gives an amplitude of $10\pm 9$ m\AA.

\section{Polarisation of metal lines in emission}\label{Sec:Emission}
Discovered by \cite{Osmer1968}, more than 300 weak metal lines in emission have been
identified by \cite{Kipper2011} in the spectrum of 89\,Her.  Kipper pointed out that these
lines: 1) present the velocity of the binary system, 2) are due to neutral metals 
with a rather low ($<$6\,eV) excitation levels, and 3) are not due to forbidden and ionic transitions.
Kipper concluded that these emission lines are formed in the circumbinary disk. 
Already \cite{Waters1993} have interpreted these weak, neutral or low-excitation-level emission lines of metals
 as a proof of the collisionally excited interaction between the stellar wind and the circumbinary disk.

\begin{figure}
\centering
\includegraphics[trim= 15.5cm 2.0cm 4.0cm 2.0cm, clip=true,width=4.1cm,angle=180]{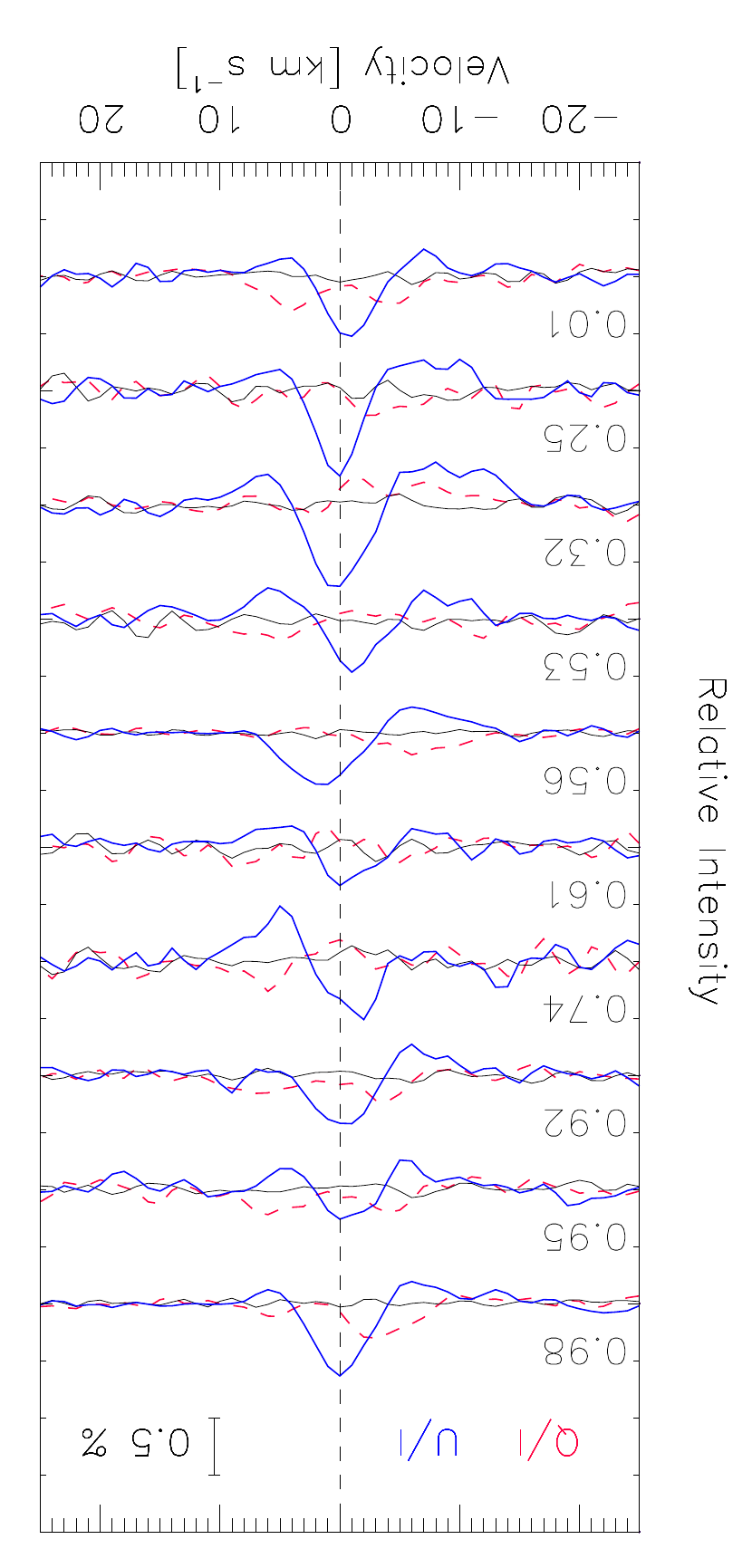}
\includegraphics[trim= 15.5cm 2.0cm 4.0cm 2.0cm, clip=true,width=4.1cm,angle=180]{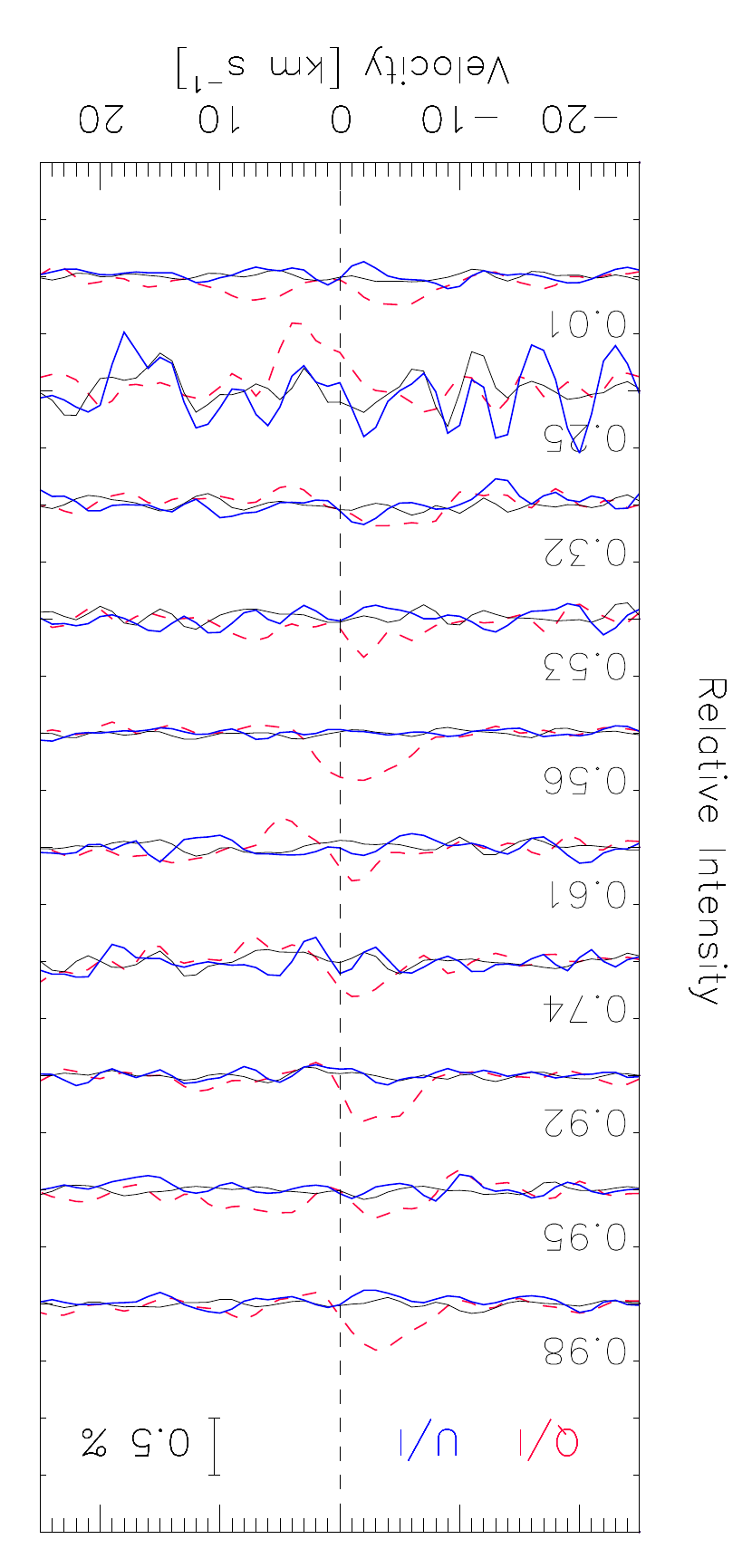}
\begin{center}\caption{\label{Fig:Emission1} For any ESPaDOnS spectrum we plot the average Stokes profiles of metal lines in emission. Phase is computed
with orbital ephemeris given in Eq.\,1 \citep{Waters1993}. Most of them, left panel,  presents the Stokes $U/I$ (solid blue) profile clearly
different than zero  and almost null Stokes $Q/I$ (dashed red) profiles. A few others are characterised by a null Stokes $U/I$ and a possible not null Stokes $Q/I$ profile (right panel). }
\end{center}
\end{figure}

We have analysed the polarisation properties of the emission metal lines crowding the spectrum of 89\,Her and found  as a general rule that
they present, left panel of Fig.\,\ref{Fig:Emission1}, an about 1\% polarisation fully stored in the Stokes $U/I$ profile.
However, few emission lines appear to show the opposite behavior, right panel of Fig.\,\ref{Fig:Emission1}, to be confirmed
with further observations.

\begin{figure}
\centering
\includegraphics[trim= 3.9cm 5cm 1.5cm 2.0cm, clip=true,width=12.0cm,angle=180]{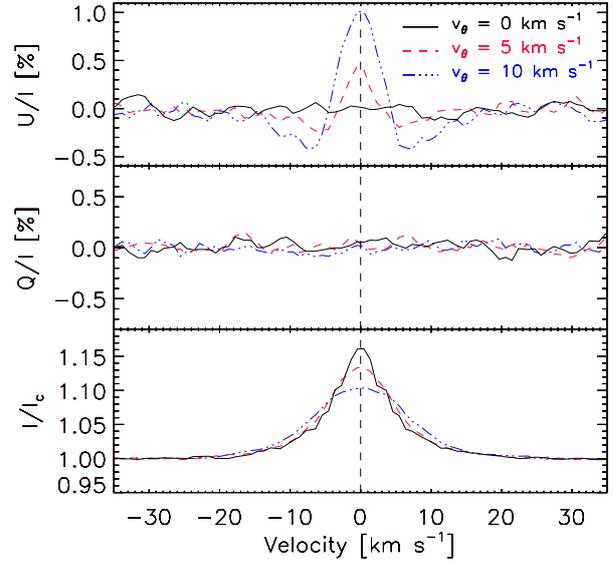}
\begin{center}\caption{\label{Fig:DiskLine}  Polarisation of a spectral line emitted from a  $30^{\circ}$ opened undisrupted disk
seen at $cos(\theta) = 0.85$. The black solid line is for a non-rotating disk, the red dashed line for a disk with rotational velocity
$\rm v_{\theta}=$ $5$  km\,s$^{-1}$ and the blue dashed-solid line for a disk with $\rm v_{\theta}=$ $10$  km\,s$^{-1}$.
To justify the null Stokes $Q$ profiles, a PA = 45$^{\circ}$ has been assumed.} \end{center}
\end{figure}

These polarised line profiles are expected for rotating disks \citep{Vink2005}. We have numerically computed polarised profiles
with the STOKES code and found that observations are justified
by an opened undisrupted disk slightly tilted towards the line of sight and rotating at $\le\,10$ km\,s$^{-1}$,  Fig.\,\ref{Fig:DiskLine}. 
A value in agreement with \cite{Buja2007} conclusion that the characteristic rotation would be $\sim 8$ km\,s$^{-1}$.
We adopted a density of scattering electrons equal to 10$^{10}$ cm$^{-3}$, an external radius of 10 AU and a PA of the circum-binary disk
 equal to $45\degr$. 
 
In the literature, an indirect evidence of free electrons in the circum-binary disk of 89\,Her is the necessity of the H$^-$ continuum opacity 
 to justify the observed spectral energy distribution \citep{Hillen2013}. We note that the mass loss \citep[$\sim 10^{-8} M_\odot yr^{-1}$,][]{Osmer1968} from the primary star, responsible 
 for the observed H$_{\alpha}$ P Cygni profile, could be a continuously source of free electrons for the low temperature circum-binary disk. 
 
 We note that the observed Stokes $U/I$ profiles are slightly variable with the orbital period. It could be that the STOKES assumption of central symmetry of the disk density does not strictly apply
because of the tidal forces induced by the orbiting components.

\section{Origin of linear polarisation of metal lines in absorption }\label{Sec:Modeling}
Spectral lines in absorption with linearly polarised profiles are not common and, as a consequence, available theoretical explanations and literature are rather limited.
{ Starting from the consideration that the polarisation of metal lines in emission is not variable with the orbital period, the phenomenon responsible for the
polarisation of metal lines in absorption is on a smaller scale than the circumbinary disk and related to the presence of the secondary.}

\subsection{Magnetic Fields}
Since \cite{wade2000} and \cite{Bagnulo2001}, linearly polarised metal lines are observed  in the spectra of magnetic chemically peculiar stars  \citep[still offers a complete and gentle
introduction to this class of stars]{Wolff1983} and it is ascribed to the \emph{Zeeman effect}. As to 89\,Her,  this possibility has been ruled out by \citet{Sabin2015}
who detected no Stokes {\it V} signal across the LSD line profile (see their figure 1) and established an upper limit of 10 G to the effective magnetic field, that is by definition  the average over the visible stellar disk of longitudinal components of the field \citep{Babcock47}. 

With CAOS, we have also performed circular spectropolarimetry of 89\,Her on  HJD56853.535 and
an upper limit of 50 G has been estimated for the effective field from the Stokes {\it V} and {\it I}  LSD profiles. 
The field has been measured with the moment technique as in \cite{Leone2004}.
In principle, a linear polarisation in spectral lines associated to a null circular polarisation is possible for a purely transverse field. However, for a large scale organised stellar magnetic field of  a rotating star even if the average of the longitudinal components across the visible stellar disk is null, Stokes $V$ profiles are different than zero. 
We  then agree with \citet{Sabin2015} conclusion and exclude the possibility we are recording the linear polarisation of Zeeman components.

\subsection{Stellar Continuum Polarisation}
{\bf Stellar Pulsations}\\
\cite{Odell1979} firstly introduced the idea that stars with non-radial pulsations show a periodically variable polarisation due to the photospheric electron-scattering opacity and then he observed the linked photopolarimetric variability with an amplitude of 0.045\%  in the star BW Vul \citep{Odell1981}.  \cite{Fabas2011} have associated the variable polarisation presented by the Balmer lines of the pulsating $\omicron$\,Ceti to the propagation of shock waves. Later, \cite{Lebre2014a}  observed a variable polarisation in the metal lines of the  Mira star $\chi$Cygni and the RV\,Tauri star R\,Sct and they related the polarisation to the global asymmetry at the photospheric level induced by pulsations. 

We  can rule out that the metal line polarisation of 89\,Her
is a consequence of pulsations simply because the observed polarisation is not variable with the 63.5 day pulsation period but rather with the orbital period, Fig.\,\ref{Fig:Scargle}.\\

\noindent{\bf Hot spots}\\
\cite{Schwarz1984} and \cite{Clarke1984}  have shown that the variable photopolarimetric data of Betelgeuse with particular reference to the TiO band can be consequence of a wavelength independent (e.g. Thomson) scattering and temporal evolution of hot spots. Extending Clarke and Schwarz's interpretation, \cite{Auriere2016} suggested that the linear polarisation observed in the spectral line of Betelgeuse is indeed an effect of the depolarisation of the continuum set to zero. 

\cite{Schwarz1984} have computed the expected polarisation of the continuum in the 400\,-\,700 nm range due to hot spots.
It appears that, whatever the temperature and extension of hot spots are, polarisation decreases with the wavelengths.
We can then conclude that in the case of 89\,Her the observed polarisation across metal lines is not due to
hot spots as these have been theoretised for Betelgeuse. We have found a polarisation degree that is independent on the wavelength, Fig.\,\ref{Fig:PolWave}.   
  
\begin{figure}
\centering
\includegraphics[trim= 7.5cm 9.0cm 4.7cm 2.2cm, clip=true,width=6.0cm,angle=180]{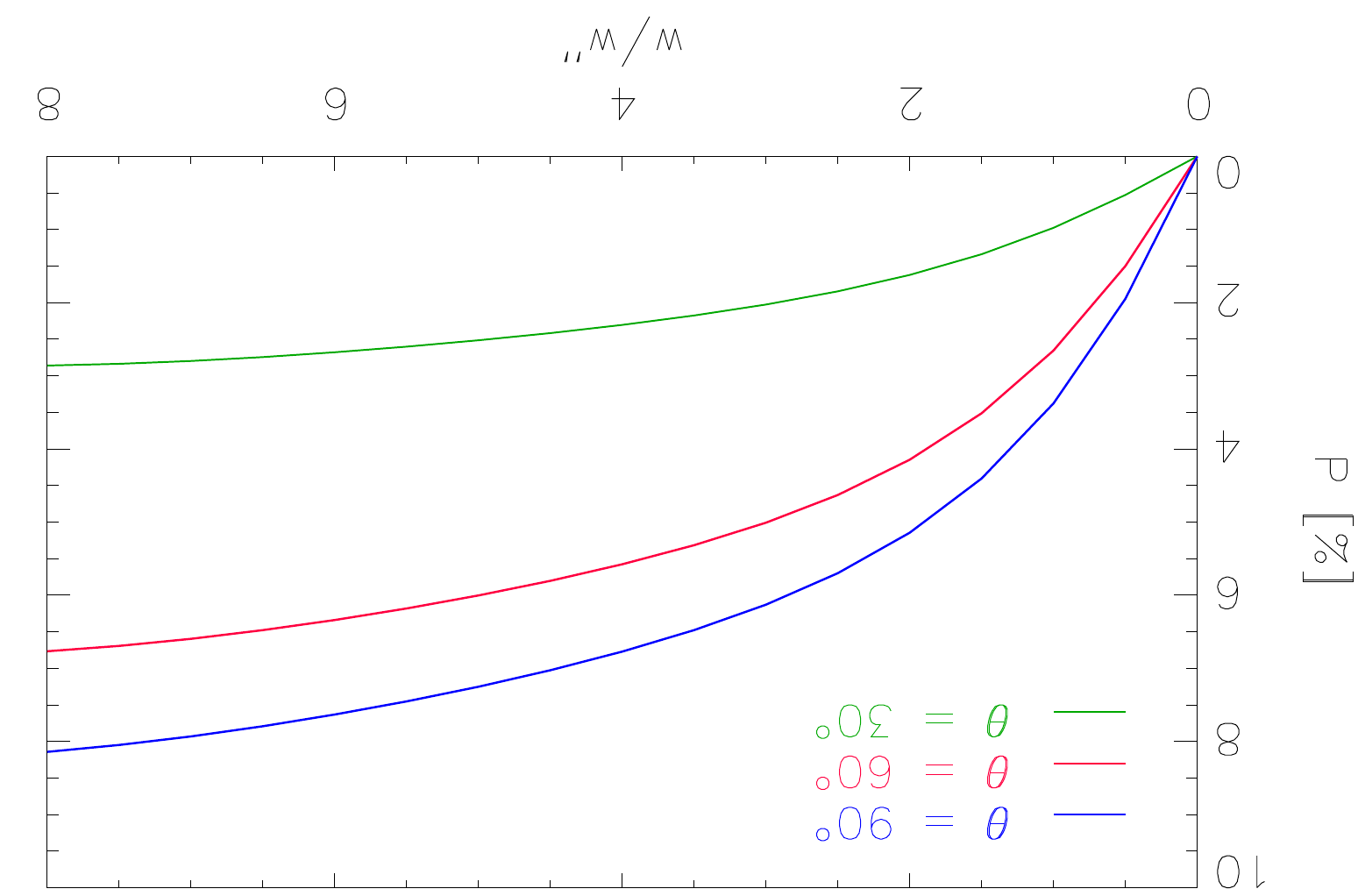}
\includegraphics[trim= 7.5cm 9.0cm 4.7cm 2.2cm, clip=true,width=6.0cm,angle=180]{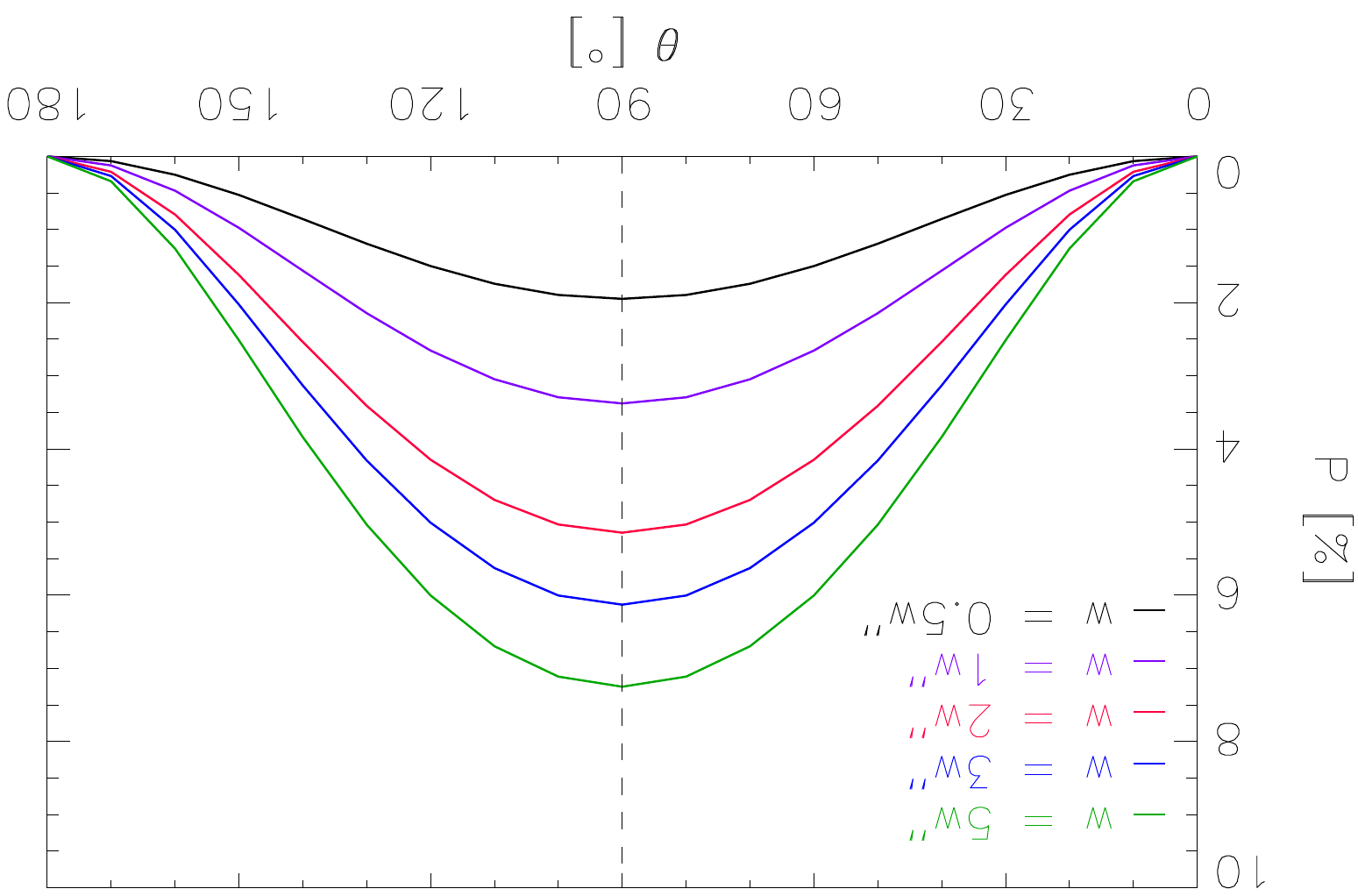}
\begin{center}\caption{\label{Fig:Pol_Theta} Spectral line polariation as a function of $\theta$ angle (bottom panel) and anisotropy factor $w$ \citep{Landi04}
normalised to the Sun value ($w\arcsec = 0.37$) at 2\arcsec (top panel).}
\end{center}
\end{figure}
  
\begin{figure}
\centering
\includegraphics[trim= 21.4cm 13cm 2.2cm 2.0cm, clip=true,width=4.cm,angle=180]{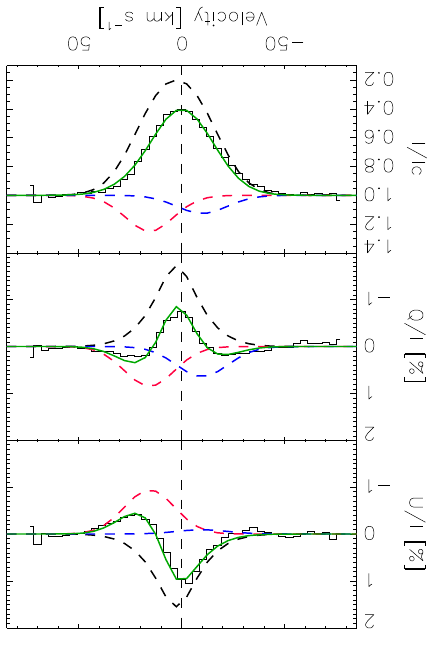}
\includegraphics[trim= 21.4cm 13cm 2.2cm 2.0cm, clip=true,width=4.cm,angle=180]{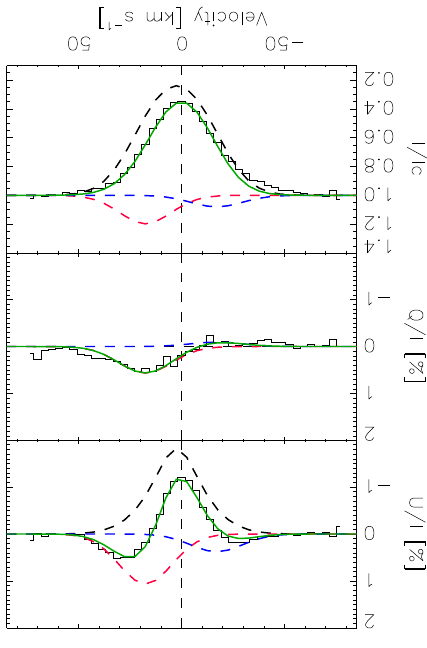}
\begin{center}\caption{\label{Fig:ResultsFit} Example of CAOS-LSD profiles (black) fitted with three components (dashed lines).
The solid (green) continuum is the sum of the three components.  From the left, spectra were acquired on HJD56816.515 and 56876.340 respectively.} \end{center}
\end{figure}

\subsection{Scattering polarisation from circumbinary electrons}\label{Sec:Scattering}
\begin{figure*}
\centering
\includegraphics[trim= 10cm 2cm 3cm 3cm, clip=true,width=5.5cm,angle=180]{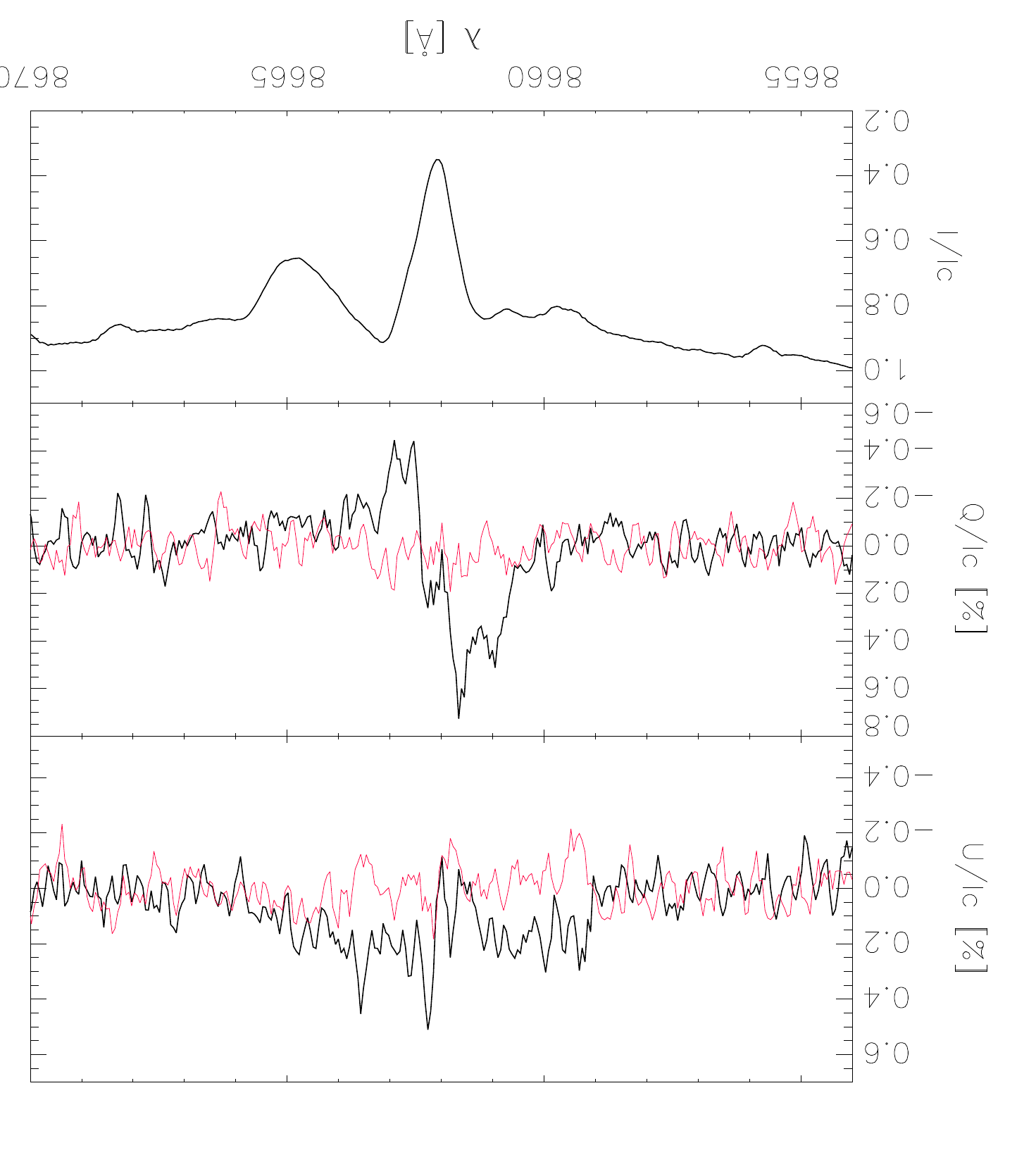}
\includegraphics[trim= 10cm 2cm 3cm 3cm, clip=true,width=5.5cm,angle=180]{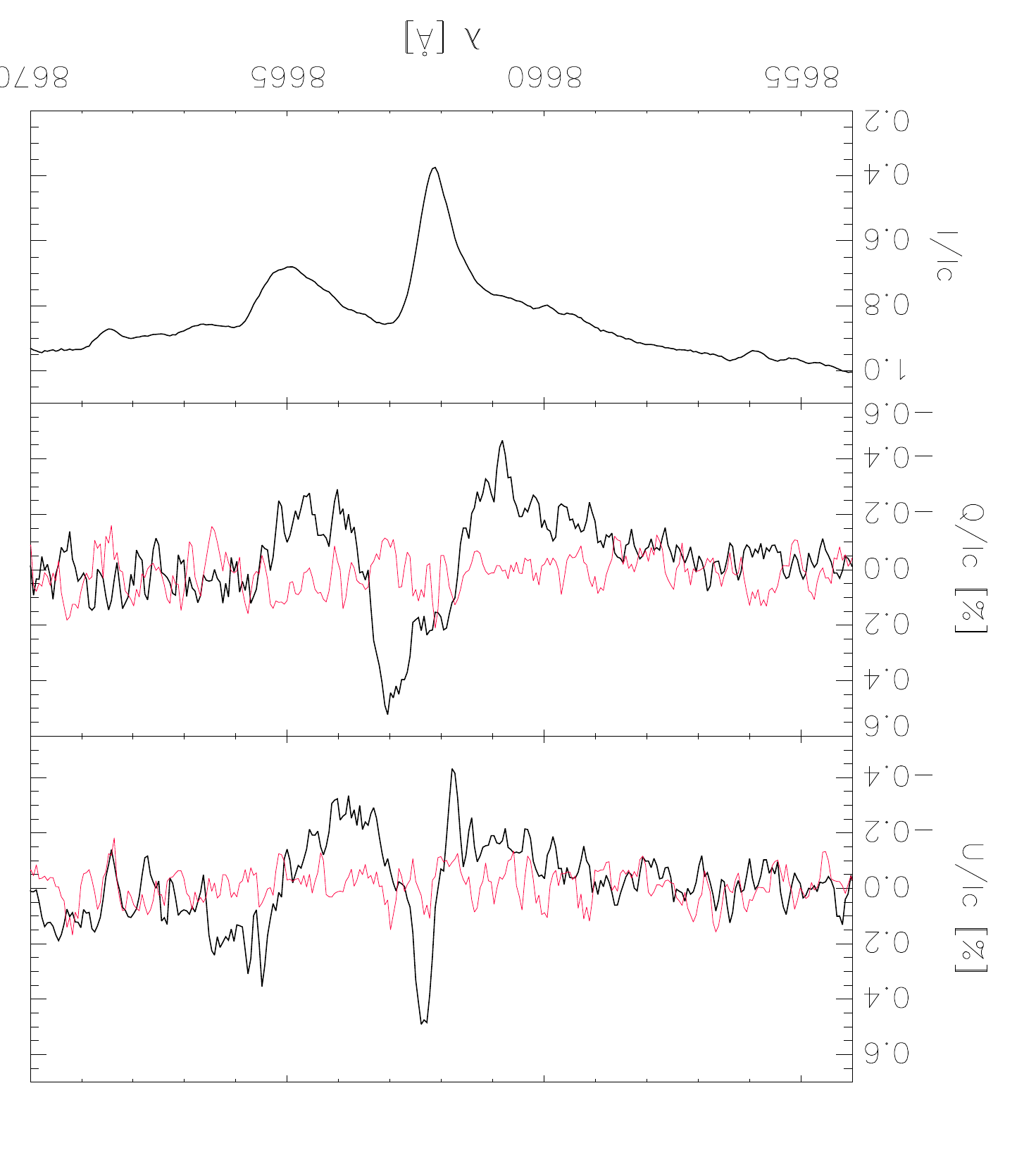}
\includegraphics[trim= 10cm 2cm 3cm 3cm, clip=true,width=5.5cm,angle=180]{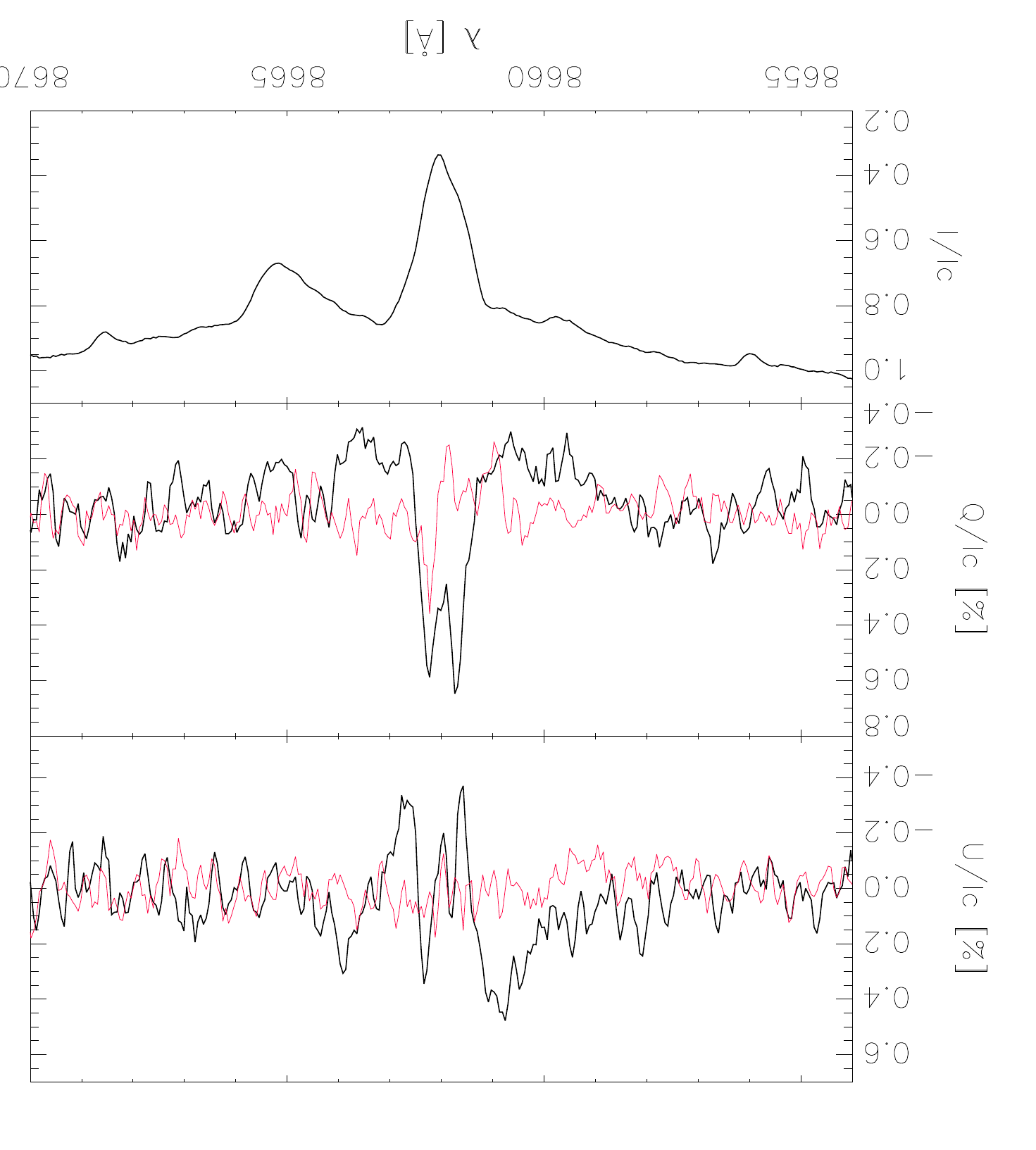}
\begin{center}\caption{\label{Fig:CaII} Stokes parameters of the Ca{\sc ii}\,8662\AA\, line as observed, from left, on HJD 53777, 53961 and 54372. According to \protect\cite{Kuhn2011} 
the very existence of the polarisation in this calcium line rules out the scattering polarisation from free electrons. It is compatible with optical pumping instead.}
\end{center}
\end{figure*}

{
In section\,\ref{Sec:Emission} we explained the polarisation of metal lines in emission with the presence of the circumbinary disk. 
Despite \cite{Hillen2013} detection of a 35-40\% optical circumstellar flux contribution to the 89\,Her luminosity,
the polarisation from free electrons of the circumbinary envelope is certainly not  at the origin of the polarisation
of metal lines in absorption. We have always clearly detected linear polarisation across the Ca{\sc ii}\,8662 \AA\, line profile (Fig.\ref{Fig:CaII}),
that, as pointed out by \cite{Kuhn2011}, cannot be due to scattering \citep{Rafa2003}.

Nevertheless, we have carried out numerical tests with the STOKES code  to understand the role of scattering polarisation from free electrons of the circumstellar environment.
Polarised line profiles at the observed 1\% level are predicted if the \cite{Hillen2013} bipolar outflow with an appropriate ($n_e \sim 10^{10} - 10^{11}$ cm$^{-3}$) electron density
is assumed. However, the ensuing 6\% polarised continuum is not compatible with \cite{Akras2017} finding of no polarisation.
In addition, we were not able to adapt the outflow properties in order to justify the polarisation variability across spectral lines with the orbital period. 
}

\subsection{Anisotropic Radiation Pumping}\label{Sec:Pumping}
\cite{Kuhn07} suggested that the linearly polarised  continuum of Herbig Ae/Be stars is due to the \emph{anisotropic radiation pumping}.
At our knowledge, this mechanism has not yet been suggested as origin of polarisation in metal spectral lines
of stars other than the Sun. 

{ The here observed linear polarisation across the Ca{\sc ii} 8662 \AA\, line profile, ruling out the scattering polarisation, is the clear evidence of optical pumping \citep{Kuhn2011}.
As} to 89\,Her, the presence of a very close \citep[0.31\,AU,][]{Buja2007} companion  justifies, at least in principle, the assumption
of a not isotropic radiation field responsible of the observed polarised metal lines be seen variable with the orbital period.
A quantitative evaluation based on the anisotropy factor $w$  \citep{Landi04}, that  is zero for an isotropic radiation field and 1 for an unidirectional radiation beam,
confirms the presence of a not isotropic field between the two components of 89\,Her. We obtain $w_{\tiny 550 nm} \sim 0.8$ 
if {\it a}) the stellar parameters by \cite{Waters1993}, who found that the primary component presents T$_{\rm eff}$ = 6500 K, $\log\,g = 1.0$ and
R = 41 R$_\odot$, while the secondary
is a 0.6\,$M_\odot$ main sequence star: that is an M0V star with  T$_{\rm eff}$ = 4045 K, $\log\,g = 4.6$  and R = 0.6 R$_\odot$ \citep{Gray2008}, and
{\it b}) the surface fluxes by \cite{Kurucz2005} and limb-darkening by \cite{Diaz1995} are adopted.
\begin{figure*}
\centering
\includegraphics[trim=2.8cm  4.cm   5.0cm  3.0cm, clip=true,width=9cm,angle=90]{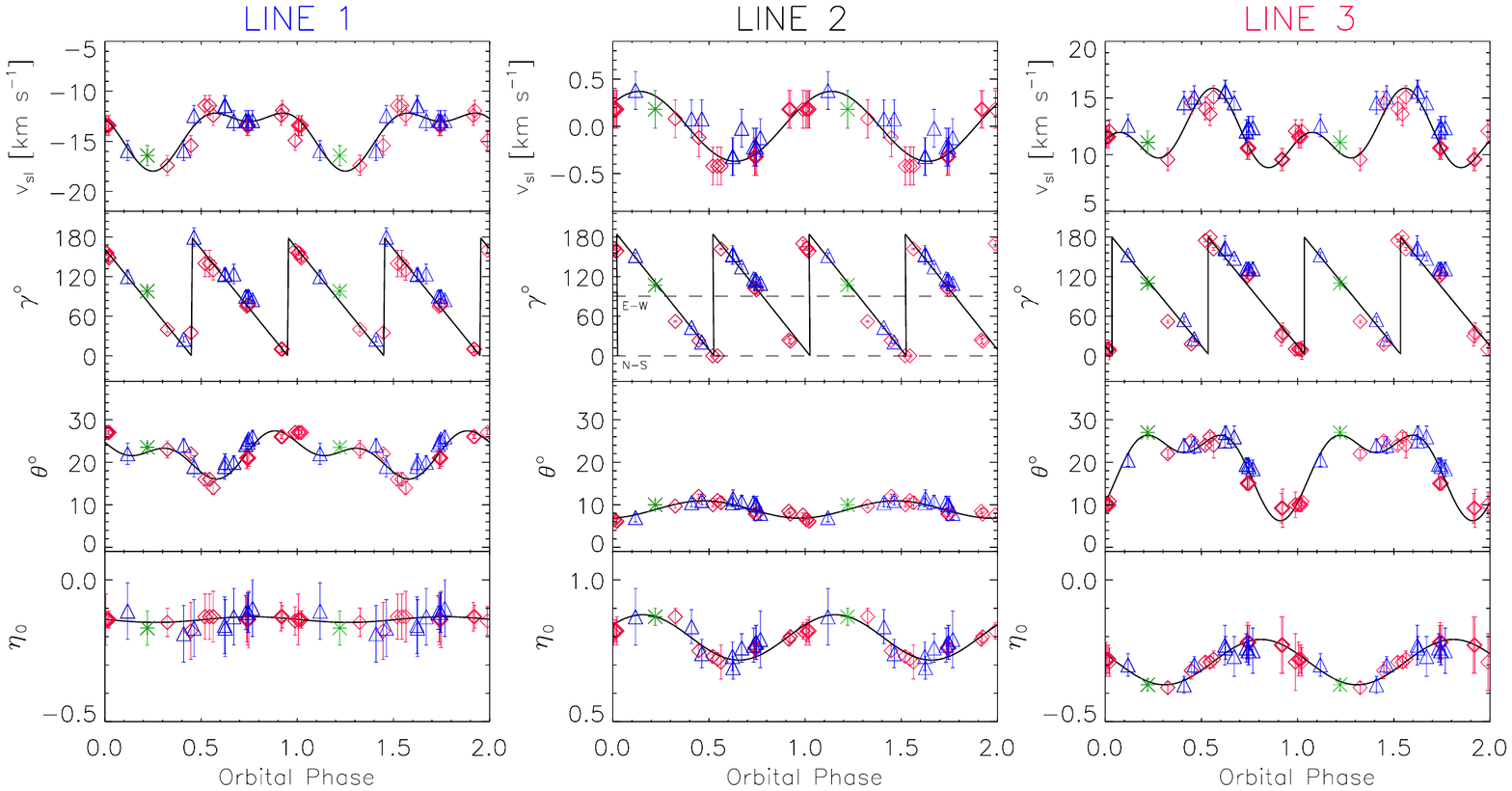}
\begin{center}\caption{\label{Fig:ResultsTime2} HAZEL parameters best matching the Stokes LSD profiles are folded with the orbital ephemeris given in Eq. \ref{Ephemeris}.
Symbols are as in Fig.\,\ref{Fig:Scargle}. The saw-tooth variation of the $\gamma$ angle, due to its definition in 0\,-\,180$\degr$ range, is indicative of a closed loop of the
polarisation vector along the orbital motion of the secondary star.  The longitude of periastron $\omega = 359.3\degr$ \citep{Waters1993} implies that the polarisation vector is North-South oriented
($\gamma = 0\degr$) in conjunctions and East-West oriented ($\gamma = 90\degr$) in quadratura. Single and double wave variations have been assumed to match the parameter changes but
for the $\gamma$ one that has been assumed linear in time.}\end{center}
\end{figure*}

To numerically explore the possibility we are really in presence of a spectral line polarisation induced by a periodically variable anisotropic radiation field due to
the orbiting secondary, we compare synthetic Fe{\sc II}\,4508.288\,\AA\, Stokes profiles computed with the HAZEL  code \citep{Asensio2008} with the observed LSD profiles.
For calculations, observed profiles have been corrected for the Doppler shift due to the orbital motion of the primary component.
Magnetic fields, that are not necessary  to induce a population imbalance \citep{Bueno1997}, have been neglected to minimise
the free parameters to the: 1) slab velocity v$_{sl}$, 2) thermal velocity v$_{th}$, 3) strength of the
Stokes $I$ line $\eta_0$, 4) angle $\theta$ between the LoS and the normal {\bf n} to the slab and 5) polarisation
angle $\gamma$ representing the azimuth of the polarisation vector with respect to the North-South direction.
Fig.\,\ref{Fig:Pol_Theta} shows how the anisotropy factor $w$ and the $\theta$ angle contribute to the polarisation level, again with the
aim to minimise the number of free parameters, the anisotropy factor has been arbitrarily fixed equal to the solar value ($w\arcsec = 0.37$) at 2\arcsec.

Three slabs are required to fit the variable Stokes $Q/I$ and $U/I$ LSD profiles of 89\,Her with HAZEL. 
A stationary feature in absorption (hereafter LINE 2 and due to SLAB 2) and two oppositely Doppler shifted features, the blue-shifted \textcolor{blue}{LINE 1}
of \textcolor{blue}{SLAB 1} and the red-shifted \textcolor{red}{LINE 3}  of \textcolor{red}{SLAB 3}. Fig.\,\ref{Fig:ResultsFit}
shows an example of the fit for two of CAOS-LSD profiles. 
With the exception of the thermal velocity v$_{th}$ that is constant in time, all fit parameters present a well defined variability with the orbital period (Fig.\,\ref{Fig:ResultsTime2}).
 
\begin{figure}
\centering
\includegraphics[trim= 0.0cm 0.0cm 2.0cm 0.0cm, clip=true,width=8.5cm,angle=0]{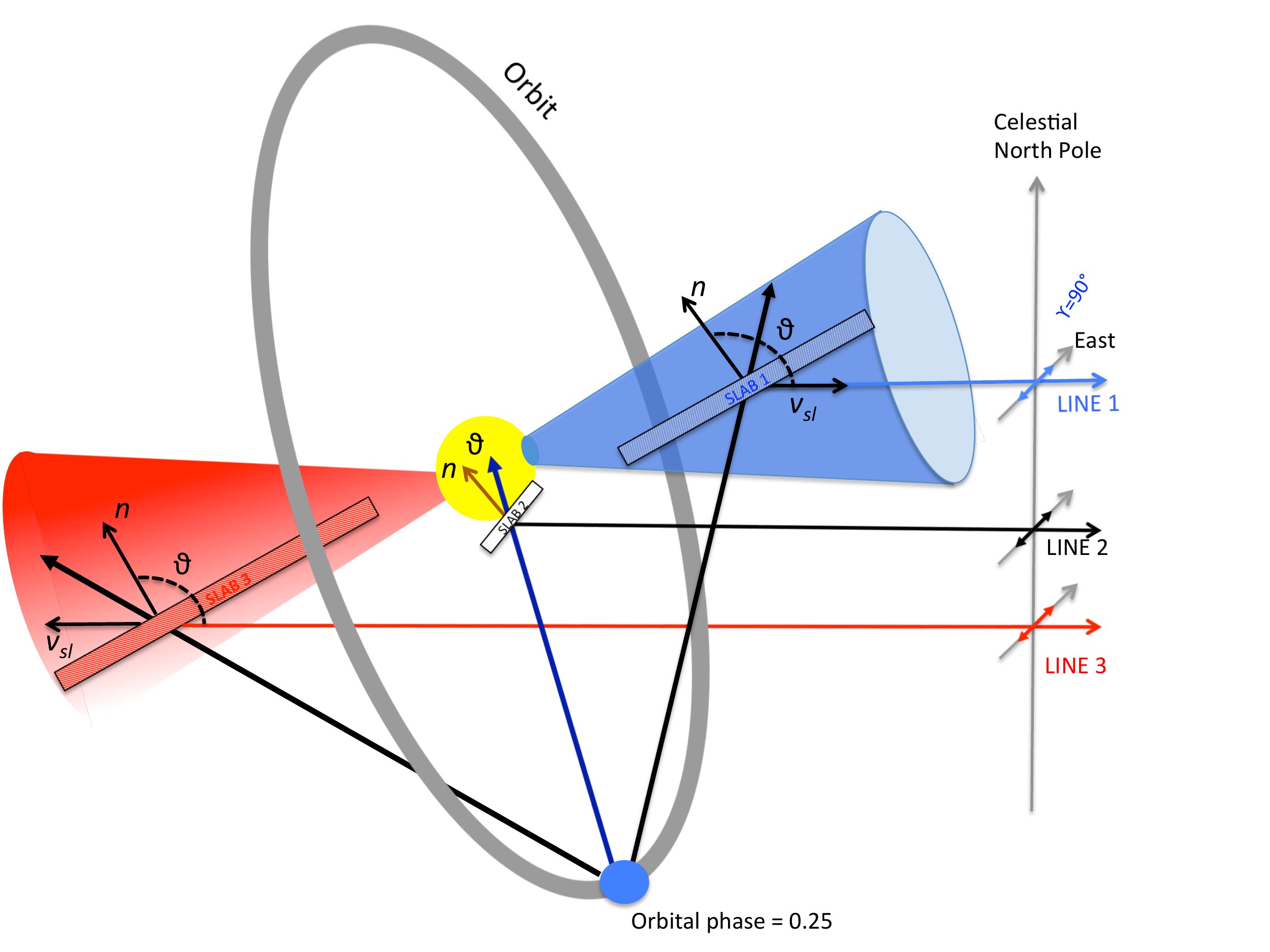}
\begin{center}\caption{\label{Fig:Sketch} A not in scale sketch of 89 Her.  The binary system presents the major axis aligned with the East-West direction  \citep{Waters1993}. An hour-glass structure is
centered on the primary component and normal to the orbital plane  \citep{Buja2007}.
In the framework of the anisotropic radiation pumping, the variable Stokes LSD profiles  (Fig.\,\ref{Fig:ResultsTime2}) can be synthetised with the HAZEL \citep{Asensio2008} code assuming three slabs:
the stationary SLAB 2 presents the {\it average} optical and physical properties of the primary component in reflecting and reprocessing
the radiation from the orbiting secondary, correspondingly the blue-shifted  \textcolor{blue}{SLAB 1} represents the jet pointing towards us and
the red-shifted \textcolor{red}{SLAB 3} represents the receding jet. $\theta$ is the angle between the normal {\bf n} to the slab and the LoS, $v_{sl}$ the velocity of the slab towards the LoS. In Section 5.3 is the complete
description and definition of fit parameters. }
\end{center}
\end{figure}

According to the adopted ephemeris (Eq.\,\ref{Ephemeris}),  the radial velocity of the primary component of the 89\,Her binary system  presents its maximum at  the orbital phase = 0
when it moves away from us and  the system is  in quadratura. Thus the less massive secondary is before the primary
(inferior conjunction) at phase = 0.25,  while the superior conjunction is at phase = 0.75. \cite{Waters1993} measured a longitude of periastron
$\omega = 359.3\degr$ orienting the main axis of the elliptical ($e = 0.189$) orbit with the East-West direction. Moreover,  89\,Her is characterised by
an hour-glass structure that is orthogonal to the orbital plane, elongated in a direction that forms
an angle of 15$\degr$ with respect to the LoS and with a Position Angle of 45$\degr$ \citep{Buja2007}. Fig.\,\ref{Fig:Sketch} shows a sketch of 89\,Her. 

We find that the overall behavior of the fit parameters of the three slabs  (Fig.\,\ref{Fig:ResultsTime2}) can be understood
within the present picture of 89\,Her, if the stationary SLAB 2 presents the {\it average} optical and physical properties of the primary component in reflecting and reprocessing
the radiation from the orbiting secondary, correspondingly the blue-shifted  \textcolor{blue}{SLAB 1} represents the jet pointing towards us and
the red-shifted \textcolor{red}{SLAB 3} represents the receding jet. In such hypothesis, it is straight to explain:
\begin{itemize}
\item the always negative \textcolor{blue}{SLAB 1} velocity  and the  always positive \textcolor{red}{SLAB 3} velocity, both of the order of the hour-glass expansion velocity equal to $\sim$6-7 \,km\,s$^{-1}$ \citep{Buja2007}. 

\item the constant thermal broadening, as it is expected for a phenomenon not due to a change of physical properties of the source, as for example because of pulsations or spots.

\item the synchronous saw-tooth variation of $\gamma$ angles for the three slabs with the orbital period. The continuous variation of $\gamma$, defined in the 0\,-\,180$\degr$ range, 
shows that the polarisation vector describes a closed loop in the sky, that is the scattering plane rotates around the LoS with the secondary star.
The orientation of the polarisation vector with the East-West direction in conjuctions and with the North-South direction in quadratura is due to the East-West
orientation of the major axis of the elliptical orbit (Fig.\,\ref{Fig:ResultsTime2} and Fig.\,\ref{Fig:Sketch}). 

\end{itemize}

A quantitative evaluation of the fit parameters and a description of their variability would require a not yet available 3-D version of  HAZEL and a detailed knowledge of 89\,Her environment.
Qualitatively, the variability of fit parameters is due to the not perfect symmetry with respect to the LoS, to the not null eccentricity of the orbit and to absorption and occultation
phenomena. For example, the weak modulation of the SLAB 2 in v$_{sl}$ and $\theta$  could be due to orbit eccentricity, or the large modulation of the \textcolor{red}{SLAB 3} $\theta$ angle with respect the \textcolor{blue}{SLAB 1}
could be a consequence of partial occultation of the receding jet by the primary star (Fig.\,\ref{Fig:Sketch}).

\section{Conclusions}\label{Sec:Discussions}
On the basis of high resolution spectropolarimetry, we have found that the metal lines of the binary system 89\,Her are polarised as in the  \emph{Second Solar Spectrum}.
These absorption metal lines with low excitation potentials  present a linear polarisation varying in time with the orbital period.

Because of the variability with the orbital period, we rule out as origin of the variability the pulsations. Moreover, we can also rule out hot spots
as possible origin the continuum depolarisation because of the not dependence of polarisation on wavelength.

{ According to \cite{Kuhn2011}, the linear polarisation we detected in the Ca{\sc ii}\,8662\AA\, line rules out 
the scattering polarisation  from free electrons  of the circumbinary environment and 
it suggests that the} origin of the \emph{Second Solar Spectrum} we observed in 89\,Her is
the anisotropic radiation field due to the secondary star. In this framework, the periodic variability of polarised profiles, particularly the closed loop described by the polarisation angle,
can be ascribed to the jets at the
basis of the hour-glass structure differently illuminated by the secondary orbiting around the primary.

A further unexpected properties of 89\,Her is the polarisation across the emission metal lines. Numerical simulations show that polarised profiles are consistent
with electron scattering in an undisrupted rotating disk. Such a disk appears to be clockwise rotated of $45\degr$ (clockwise) from the North-South direction as it was observed by interferometric observations.

 If it is possible to generalise the result of our observational campaign of 89\,Her, we conclude that the inner regions of  stellar enevlopes
 can be probed via high resolution spectropolarimetry and that this appears the only diagnostics for stars still too far for interferometric studies.

\section*{Acknowledgements}
This research used the facilities of the Canadian Astronomy Data Centre operated by the National Research Council of Canada with the support of the Canadian Space Agency.
Based on observations made with the Italian Telescopio Nazionale Galileo (TNG) operated on the island of La Palma by the Fundación Galileo Galilei of the INAF
(Istituto Nazionale di Astrofisica) at the Spanish Observatorio del Roque de los Muchachos of the Instituto de Astrofisica de Canarias.

\bibliographystyle{mnras}
\bibliography{SecondSolar} 

\bsp	
\label{lastpage}
\end{document}

%% file: ObservingLog.tex
\begin{table}
\caption{\label{tab:ObsLog} Logbook of observations.  The achieved Signal-to-Noise (S/N) is determined from \emph{null} LSD spectra.}
\begin{tabular}{crccrc}
\hline
HJD  &  \multicolumn{1}{c}{S/N}  & Instr.& HJD  &  \multicolumn{1}{c}{S/N} & Instr.\\
2400000+  &   & & 2400000+  &   & \\
\hline
53604.753 & 11300 & ESP  & 55102.832 & 5200  & ESP  \\
53775.115 & 10500 & ESP  & 55109.823 & 12400 & ESP  \\
53777.096 & 19400 & ESP  & 56816.515 & 3000  & CAOS\\   
53961.777 & 21800 & ESP  & 56862.456 & 2300 & CAOS \\  
54372.821 & 23900 & ESP  & 56863.347 & 2400  & CAOS\\   
54877.099 & 4300  & ESP  & 56876.340 & 3500  & CAOS\\   
54878.132 & 10700 & ESP  & 56894.309 & 4200  & CAOS\\   
54880.076 & 14700 & ESP  & 56897.301 & 4100  & CAOS\\   
54954.125 & 12900 & ESP  & 56898.274 & 4300  & CAOS\\   
54955.849 & 10600 & ESP  & 56904.313 & 3700  & CAOS \\  
54958.121 & 14000 & ESP  & 57582.248 & 3700  & CAOS \\  
54959.874 & 8000  & ESP  & 57666.248 & 2900  & CAOS  \\ 
55081.874 & 3000  & ESP  & 57900.100 & 4300  & HANPO \\ 
\hline
\end{tabular}
\end{table}

%% file: SecondSolar.bbl
\begin{thebibliography}{}
\makeatletter
\relax
\def\mn@urlcharsother{\let\do\@makeother \do\$\do\&\do\#\do\^\do\_\do\%\do\~}
\def\mn@doi{\begingroup\mn@urlcharsother \@ifnextchar [ {\mn@doi@}
  {\mn@doi@[]}}
\def\mn@doi@[#1]#2{\def\@tempa{#1}\ifx\@tempa\@empty \href
  {http://dx.doi.org/#2} {doi:#2}\else \href {http://dx.doi.org/#2} {#1}\fi
  \endgroup}
\def\mn@eprint#1#2{\mn@eprint@#1:#2::\@nil}
\def\mn@eprint@arXiv#1{\href {http://arxiv.org/abs/#1} {{\tt arXiv:#1}}}
\def\mn@eprint@dblp#1{\href {http://dblp.uni-trier.de/rec/bibtex/#1.xml}
  {dblp:#1}}
\def\mn@eprint@#1:#2:#3:#4\@nil{\def\@tempa {#1}\def\@tempb {#2}\def\@tempc
  {#3}\ifx \@tempc \@empty \let \@tempc \@tempb \let \@tempb \@tempa \fi \ifx
  \@tempb \@empty \def\@tempb {arXiv}\fi \@ifundefined
  {mn@eprint@\@tempb}{\@tempb:\@tempc}{\expandafter \expandafter \csname
  mn@eprint@\@tempb\endcsname \expandafter{\@tempc}}}

\bibitem[\protect\citeauthoryear{{Akras}, {Ram{\'{\i}}rez V{\'e}lez},
  {Nanouris}, {Ramos-Larios}, {L{\'o}pez}, {Hiriart}  \& {Panoglou}}{{Akras}
  et~al.}{2017}]{Akras2017}
{Akras} S.,  {Ram{\'{\i}}rez V{\'e}lez} J.~C.,  {Nanouris} N.,  {Ramos-Larios}
  G.,  {L{\'o}pez} J.~M.,  {Hiriart} D.,   {Panoglou} D.,  2017, \mn@doi
  [\mnras] {10.1093/mnras/stw3046}, \href
  {http://adsabs.harvard.edu/abs/2017MNRAS.466.2948A} {466, 2948}

\bibitem[\protect\citeauthoryear{{Arellano Ferro}}{{Arellano
  Ferro}}{1984}]{Ferro1984}
{Arellano Ferro} A.,  1984, \mn@doi [\pasp] {10.1086/131397}, \href
  {http://adsabs.harvard.edu/abs/1984PASP...96..641A} {96, 641}

\bibitem[\protect\citeauthoryear{{Asensio Ramos}, {Trujillo Bueno}  \& {Landi
  Degl'Innocenti}}{{Asensio Ramos} et~al.}{2008}]{Asensio2008}
{Asensio Ramos} A.,  {Trujillo Bueno} J.,   {Landi Degl'Innocenti} E.,  2008,
  \mn@doi [\apj] {10.1086/589433}, \href
  {http://adsabs.harvard.edu/abs/2008ApJ...683..542A} {683, 542}

\bibitem[\protect\citeauthoryear{{Asensio Ramos}, {Mart{\'{\i}}nez
  Gonz{\'a}lez}, {Manso Sainz}, {Corradi}  \& {Leone}}{{Asensio Ramos}
  et~al.}{2014}]{Asensio2014}
{Asensio Ramos} A.,  {Mart{\'{\i}}nez Gonz{\'a}lez} M.~J.,  {Manso Sainz} R.,
  {Corradi} R.~L.~M.,   {Leone} F.,  2014, \mn@doi [\apj]
  {10.1088/0004-637X/787/2/111}, \href
  {http://adsabs.harvard.edu/abs/2014ApJ...787..111A} {787, 111}

\bibitem[\protect\citeauthoryear{{Auri{\`e}re} et~al.,}{{Auri{\`e}re}
  et~al.}{2016}]{Auriere2016}
{Auri{\`e}re} M.,  et~al., 2016, \mn@doi [\aap] {10.1051/0004-6361/201628077},
  \href {http://adsabs.harvard.edu/abs/2016A%26A...591A.119A} {591, A119}

\bibitem[\protect\citeauthoryear{{Babcock}}{{Babcock}}{1947}]{Babcock47}
{Babcock} H.~W.,  1947, \mn@doi [ApJ] {10.1086/144887}, 105, 105

\bibitem[\protect\citeauthoryear{{Bagnulo}, {Wade}, {Donati}, {Landstreet},
  {Leone}, {Monin}  \& {Stift}}{{Bagnulo} et~al.}{2001}]{Bagnulo2001}
{Bagnulo} S.,  {Wade} G.~A.,  {Donati} J.-F.,  {Landstreet} J.~D.,  {Leone} F.,
   {Monin} D.~N.,   {Stift} M.~J.,  2001, \mn@doi [\aap]
  {10.1051/0004-6361:20010101}, \href
  {http://adsabs.harvard.edu/abs/2001A%26A...369..889B} {369, 889}

\bibitem[\protect\citeauthoryear{{Balick} \& {Frank}}{{Balick} \&
  {Frank}}{2002}]{Balick2002}
{Balick} B.,  {Frank} A.,  2002, \mn@doi [\araa]
  {10.1146/annurev.astro.40.060401.093849}, \href
  {http://adsabs.harvard.edu/abs/2002ARA%26A..40..439B} {40, 439}

\bibitem[\protect\citeauthoryear{{Bujarrabal}, {van Winckel}, {Neri},
  {Alcolea}, {Castro-Carrizo}  \& {Deroo}}{{Bujarrabal}
  et~al.}{2007}]{Buja2007}
{Bujarrabal} V.,  {van Winckel} H.,  {Neri} R.,  {Alcolea} J.,
  {Castro-Carrizo} A.,   {Deroo} P.,  2007, \mn@doi [\aap]
  {10.1051/0004-6361:20066969}, \href
  {http://adsabs.harvard.edu/abs/2007A%26A...468L..45B} {468, L45}

\bibitem[\protect\citeauthoryear{{Clarke} \& {Schwarz}}{{Clarke} \&
  {Schwarz}}{1984}]{Clarke1984}
{Clarke} D.,  {Schwarz} H.~E.,  1984, \aap, \href
  {http://esoads.eso.org/abs/1984A%26A...132..375C} {132, 375}

\bibitem[\protect\citeauthoryear{{Diaz-Cordoves}, {Claret}  \&
  {Gimenez}}{{Diaz-Cordoves} et~al.}{1995}]{Diaz1995}
{Diaz-Cordoves} J.,  {Claret} A.,   {Gimenez} A.,  1995, \aaps, \href
  {http://adsabs.harvard.edu/abs/1995A%26AS..110..329D} {110, 329}

\bibitem[\protect\citeauthoryear{{Donati}, {Semel}, {Carter}, {Rees}  \&
  {Collier Cameron}}{{Donati} et~al.}{1997}]{Donati1997}
{Donati} J.-F.,  {Semel} M.,  {Carter} B.~D.,  {Rees} D.~E.,   {Collier
  Cameron} A.,  1997, \mn@doi [\mnras] {10.1093/mnras/291.4.658}, \href
  {http://adsabs.harvard.edu/abs/1997MNRAS.291..658D} {291, 658}

\bibitem[\protect\citeauthoryear{{Donati}, {Catala}, {Landstreet}  \&
  {Petit}}{{Donati} et~al.}{2006}]{Donati2006}
{Donati} J.-F.,  {Catala} C.,  {Landstreet} J.~D.,   {Petit} P.,  2006, in
  {Casini} R.,  {Lites} B.~W.,  eds,  Astronomical Society of the Pacific
  Conference Series Vol. 358, Astronomical Society of the Pacific Conference
  Series. p.~362

\bibitem[\protect\citeauthoryear{{Fabas}, {L{\`e}bre}  \& {Gillet}}{{Fabas}
  et~al.}{2011}]{Fabas2011}
{Fabas} N.,  {L{\`e}bre} A.,   {Gillet} D.,  2011, \mn@doi [\aap]
  {10.1051/0004-6361/201117748}, \href
  {http://adsabs.harvard.edu/abs/2011A%26A...535A..12F} {535, A12}

\bibitem[\protect\citeauthoryear{{Gray}}{{Gray}}{2008}]{Gray2008}
{Gray} D.~F.,  2008, {The Observation and Analysis of Stellar Photospheres}

\bibitem[\protect\citeauthoryear{{Gray} \& {Garrison}}{{Gray} \&
  {Garrison}}{1989}]{Gray1989}
{Gray} R.~O.,  {Garrison} R.~F.,  1989, \mn@doi [\apjs] {10.1086/191315}, \href
  {http://adsabs.harvard.edu/abs/1989ApJS...69..301G} {69, 301}

\bibitem[\protect\citeauthoryear{{Harrington} \& {Kuhn}}{{Harrington} \&
  {Kuhn}}{2009a}]{Harrington2009bis}
{Harrington} D.~M.,  {Kuhn} J.~R.,  2009a, \mn@doi [\apjs]
  {10.1088/0067-0049/180/1/138}, \href
  {http://adsabs.harvard.edu/abs/2009ApJS..180..138H} {180, 138}

\bibitem[\protect\citeauthoryear{{Harrington} \& {Kuhn}}{{Harrington} \&
  {Kuhn}}{2009b}]{Harrington2009}
{Harrington} D.~M.,  {Kuhn} J.~R.,  2009b, \mn@doi [\apj]
  {10.1088/0004-637X/695/1/238}, \href
  {http://adsabs.harvard.edu/abs/2009ApJ...695..238H} {695, 238}

\bibitem[\protect\citeauthoryear{{Hillen} et~al.,}{{Hillen}
  et~al.}{2013}]{Hillen2013}
{Hillen} M.,  et~al., 2013, \mn@doi [\aap] {10.1051/0004-6361/201321616}, \href
  {http://adsabs.harvard.edu/abs/2013A%26A...559A.111H} {559, A111}

\bibitem[\protect\citeauthoryear{{Jordan}, {Bagnulo}, {Werner}  \&
  {O'Toole}}{{Jordan} et~al.}{2012}]{Jordan2012}
{Jordan} S.,  {Bagnulo} S.,  {Werner} K.,   {O'Toole} S.~J.,  2012, \mn@doi
  [\aap] {10.1051/0004-6361/201219175}, \href
  {http://adsabs.harvard.edu/abs/2012A%26A...542A..64J} {542, A64}

\bibitem[\protect\citeauthoryear{{Kipper}}{{Kipper}}{2011}]{Kipper2011}
{Kipper} T.,  2011, Baltic Astronomy, \href
  {http://adsabs.harvard.edu/abs/2011BaltA..20...65K} {20, 65}

\bibitem[\protect\citeauthoryear{{Kruszewski}, {Gehrels}  \&
  {Serkowski}}{{Kruszewski} et~al.}{1968}]{1968AJ.....73..677K}
{Kruszewski} A.,  {Gehrels} T.,   {Serkowski} K.,  1968, \mn@doi [\aj]
  {10.1086/110678}, \href {http://adsabs.harvard.edu/abs/1968AJ.....73..677K}
  {73, 677}

\bibitem[\protect\citeauthoryear{{Kuhn}, {Berdyugina}, {Fluri}, {Harrington}
  \& {Stenflo}}{{Kuhn} et~al.}{2007}]{Kuhn07}
{Kuhn} J.~R.,  {Berdyugina} S.~V.,  {Fluri} D.~M.,  {Harrington} D.~M.,
  {Stenflo} J.~O.,  2007, \mn@doi [\apjl] {10.1086/522425}, \href
  {http://adsabs.harvard.edu/abs/2007ApJ...668L..63K} {668, L63}

\bibitem[\protect\citeauthoryear{{Kuhn}, {Geiss}  \& {Harrington}}{{Kuhn}
  et~al.}{2011}]{Kuhn2011}
{Kuhn} J.~R.,  {Geiss} B.,   {Harrington} D.~M.,  2011, in {Kuhn} J.~R.,
  {Harrington} D.~M.,  {Lin} H.,  {Berdyugina} S.~V.,  {Trujillo-Bueno} J.,
  {Keil} S.~L.,   {Rimmele} T.,  eds,  Astronomical Society of the Pacific
  Conference Series Vol. 437, Solar Polarization 6. p.~245 (\mn@eprint {arXiv}
  {1010.0705})

\bibitem[\protect\citeauthoryear{{Kurucz}}{{Kurucz}}{2005}]{Kurucz2005}
{Kurucz} R.~L.,  2005, Memorie della Societa Astronomica Italiana Supplementi,
  \href {http://adsabs.harvard.edu/abs/2005MSAIS...8...14K} {8, 14}

\bibitem[\protect\citeauthoryear{{Landi Degl'Innocenti} \& {Landolfi}}{{Landi
  Degl'Innocenti} \& {Landolfi}}{2004}]{Landi04}
{Landi Degl'Innocenti} E.,  {Landolfi} M.,  2004, {Polarization in Spectral
  Lines}.
 Astrophysics and Space Science Library Vol. 307,
  \mn@doi{10.1007/978-1-4020-2415-3, }

\bibitem[\protect\citeauthoryear{{L{\`e}bre}, {Fabas}  \& {Gillet}}{{L{\`e}bre}
  et~al.}{2011}]{Lebre2011}
{L{\`e}bre} A.,  {Fabas} N.,   {Gillet} D.,  2011, in {Johns-Krull} C.,
  {Browning} M.~K.,   {West} A.~A.,  eds,  Astronomical Society of the Pacific
  Conference Series Vol. 448, 16th Cambridge Workshop on Cool Stars, Stellar
  Systems, and the Sun. p.~999

\bibitem[\protect\citeauthoryear{{L{\`e}bre}, {Auri{\`e}re}, {Fabas}, {Gillet},
  {Herpin}, {Petit}  \& {Konstantinova-Antova}}{{L{\`e}bre}
  et~al.}{2014}]{Lebre2014a}
{L{\`e}bre} A.,  {Auri{\`e}re} M.,  {Fabas} N.,  {Gillet} D.,  {Herpin} F.,
  {Petit} P.,   {Konstantinova-Antova} R.,  2014, in {Petit} P.,  {Jardine} M.,
    {Spruit} H.~C.,  eds,  IAU Symposium Vol. 302, Magnetic Fields throughout
  Stellar Evolution. pp 385--388, \mn@doi{10.1017/S1743921314002579}

\bibitem[\protect\citeauthoryear{{Leone} \& {Catanzaro}}{{Leone} \&
  {Catanzaro}}{2004}]{Leone2004}
{Leone} F.,  {Catanzaro} G.,  2004, \mn@doi [\aap]
  {10.1051/0004-6361:20047180}, \href
  {http://adsabs.harvard.edu/abs/2004A%26A...425..271L} {425, 271}

\bibitem[\protect\citeauthoryear{{Leone}, {Mart{\'{\i}}nez Gonz{\'a}lez},
  {Corradi}, {Privitera}  \& {Manso Sainz}}{{Leone} et~al.}{2011}]{Leone2011}
{Leone} F.,  {Mart{\'{\i}}nez Gonz{\'a}lez} M.~J.,  {Corradi} R.~L.~M.,
  {Privitera} G.,   {Manso Sainz} R.,  2011, \mn@doi [\apjl]
  {10.1088/2041-8205/731/2/L33}, \href
  {http://adsabs.harvard.edu/abs/2011ApJ...731L..33L} {731, L33}

\bibitem[\protect\citeauthoryear{{Leone}, {Corradi}, {Mart{\'{\i}}nez
  Gonz{\'a}lez}, {Asensio Ramos}  \& {Manso Sainz}}{{Leone}
  et~al.}{2014}]{Leone2014}
{Leone} F.,  {Corradi} R.~L.~M.,  {Mart{\'{\i}}nez Gonz{\'a}lez} M.~J.,
  {Asensio Ramos} A.,   {Manso Sainz} R.,  2014, \mn@doi [\aap]
  {10.1051/0004-6361/201322753}, \href
  {http://adsabs.harvard.edu/abs/2014A%26A...563A..43L} {563, A43}

\bibitem[\protect\citeauthoryear{{Leone} et~al.,}{{Leone}
  et~al.}{2016a}]{Leone2016}
{Leone} F.,  et~al., 2016a, \mn@doi [\aj] {10.3847/0004-6256/151/5/116}, \href
  {http://adsabs.harvard.edu/abs/2016AJ....151..116L} {151, 116}

\bibitem[\protect\citeauthoryear{{Leone} et~al.,}{{Leone}
  et~al.}{2016b}]{LeoneH2016}
{Leone} F.,  et~al., 2016b, in Ground-based and Airborne Instrumentation for
  Astronomy VI. p. 99087K, \mn@doi{10.1117/12.2231781}

\bibitem[\protect\citeauthoryear{{Manso Sainz} \& {Trujillo Bueno}}{{Manso
  Sainz} \& {Trujillo Bueno}}{2003}]{Rafa2003}
{Manso Sainz} R.,  {Trujillo Bueno} J.,  2003, \mn@doi [Physical Review
  Letters] {10.1103/PhysRevLett.91.111102}, \href
  {http://adsabs.harvard.edu/abs/2003PhRvL..91k1102M} {91, 111102}

\bibitem[\protect\citeauthoryear{{McLean} \& {Coyne}}{{McLean} \&
  {Coyne}}{1978}]{McLean1978}
{McLean} I.~S.,  {Coyne} G.~V.,  1978, \mn@doi [\apjl] {10.1086/182851}, \href
  {http://adsabs.harvard.edu/abs/1978ApJ...226L.145M} {226, L145}

\bibitem[\protect\citeauthoryear{{Odell}}{{Odell}}{1979}]{Odell1979}
{Odell} A.~P.,  1979, \mn@doi [\pasp] {10.1086/130492}, \href
  {http://adsabs.harvard.edu/abs/1979PASP...91..326O} {91, 326}

\bibitem[\protect\citeauthoryear{{Odell}}{{Odell}}{1981}]{Odell1981}
{Odell} A.~P.,  1981, \mn@doi [\apjl] {10.1086/183557}, \href
  {http://adsabs.harvard.edu/abs/1981ApJ...246L..77O} {246, L77}

\bibitem[\protect\citeauthoryear{{Osmer}}{{Osmer}}{1968}]{Osmer1968}
{Osmer} P.,  1968, \mn@doi [\pasp] {10.1086/128687}, \href
  {http://adsabs.harvard.edu/abs/1968PASP...80..563O} {80, 563}

\bibitem[\protect\citeauthoryear{{Roberts}, {Lehar}  \& {Dreher}}{{Roberts}
  et~al.}{1987}]{Roberts1987}
{Roberts} D.~H.,  {Lehar} J.,   {Dreher} J.~W.,  1987, \mn@doi [\aj]
  {10.1086/114383}, \href {http://adsabs.harvard.edu/abs/1987AJ.....93..968R}
  {93, 968}

\bibitem[\protect\citeauthoryear{{Sabin}, {Wade}  \& {L{\`e}bre}}{{Sabin}
  et~al.}{2015}]{Sabin2015}
{Sabin} L.,  {Wade} G.~A.,   {L{\`e}bre} A.,  2015, \mn@doi [\mnras]
  {10.1093/mnras/stu2227}, \href
  {http://adsabs.harvard.edu/abs/2015MNRAS.446.1988S} {446, 1988}

\bibitem[\protect\citeauthoryear{{Scargle}}{{Scargle}}{1982}]{Scargle1982}
{Scargle} J.~D.,  1982, \mn@doi [\apj] {10.1086/160554}, \href
  {http://adsabs.harvard.edu/abs/1982ApJ...263..835S} {263, 835}

\bibitem[\protect\citeauthoryear{{Schwarz} \& {Clarke}}{{Schwarz} \&
  {Clarke}}{1984}]{Schwarz1984}
{Schwarz} H.~E.,  {Clarke} D.,  1984, \aap, \href
  {http://esoads.eso.org/abs/1984A%26A...132..370S} {132, 370}

\bibitem[\protect\citeauthoryear{{Stenflo}}{{Stenflo}}{1982}]{Stenflo1982}
{Stenflo} J.~O.,  1982, \mn@doi [\solphys] {10.1007/BF00147969}, \href
  {http://adsabs.harvard.edu/abs/1982SoPh...80..209S} {80, 209}

\bibitem[\protect\citeauthoryear{{Tessore}, {L{\`e}bre}  \& {Morin}}{{Tessore}
  et~al.}{2015}]{Lebre2015}
{Tessore} B.,  {L{\`e}bre} A.,   {Morin} J.,  2015, in {Martins} F.,
  {Boissier} S.,  {Buat} V.,  {Cambr{\'e}sy} L.,   {Petit} P.,  eds, SF2A-2015:
  Proceedings of the Annual meeting of the French Society of Astronomy and
  Astrophysics. pp 429--433

\bibitem[\protect\citeauthoryear{{Trujillo Bueno} \& {Landi
  Degl'Innocenti}}{{Trujillo Bueno} \& {Landi
  Degl'Innocenti}}{1997}]{Bueno1997}
{Trujillo Bueno} J.,  {Landi Degl'Innocenti} E.,  1997, \mn@doi [\apjl]
  {10.1086/310713}, \href {http://adsabs.harvard.edu/abs/1997ApJ...482L.183T}
  {482, L183}

\bibitem[\protect\citeauthoryear{{Vink}, {Harries}  \& {Drew}}{{Vink}
  et~al.}{2005}]{Vink2005}
{Vink} J.~S.,  {Harries} T.~J.,   {Drew} J.~E.,  2005, \mn@doi [\aap]
  {10.1051/0004-6361:20041463}, \href
  {http://adsabs.harvard.edu/abs/2005A%26A...430..213V} {430, 213}

\bibitem[\protect\citeauthoryear{{Wade}, {Donati}, {Landstreet}  \&
  {Shorlin}}{{Wade} et~al.}{2000}]{wade2000}
{Wade} G.~A.,  {Donati} J.-F.,  {Landstreet} J.~D.,   {Shorlin} S.~L.~S.,
  2000, \mn@doi [\mnras] {10.1046/j.1365-8711.2000.03273.x}, \href
  {http://adsabs.harvard.edu/abs/2000MNRAS.313..823W} {313, 823}

\bibitem[\protect\citeauthoryear{{Waters}, {Waelkens}, {Mayor}  \&
  {Trams}}{{Waters} et~al.}{1993}]{Waters1993}
{Waters} L.~B.~F.~M.,  {Waelkens} C.,  {Mayor} M.,   {Trams} N.~R.,  1993,
  \aap, \href {http://adsabs.harvard.edu/abs/1993A%26A...269..242W} {269, 242}

\bibitem[\protect\citeauthoryear{{Wolff}}{{Wolff}}{1983}]{Wolff1983}
{Wolff} S.~C.,  1983, {The A-type stars: problems and perspectives.}

\makeatother
\end{thebibliography}
